\documentclass[aps, prd, twocolumn, nofootinbib, altaffilletter, superscriptaddress]{revtex4-2}

\usepackage{amsmath}
\usepackage{amsfonts}
\usepackage{amssymb}

\usepackage{xcolor}
\usepackage[colorlinks=true, linkcolor=blue, citecolor=blue, urlcolor=blue]{hyperref}

\usepackage{tikz}

\usepackage{verbatim}

\newcommand{\result}[1]{#1}

\newcommand{\externalresult}[1]{#1}

\begin{document}

\title{
Compact Binary Coalescence Sensitivity Estimates with Injection Campaigns during the LIGO-Virgo-KAGRA Collaborations' Fourth Observing Run
}


\author{Reed Essick}
\email{essick@cita.utoronto.ca}
\affiliation{Canadian Institute for Theoretical Astrophysics, University of Toronto, 60 St. George Street, Toronto, ON M5S 3H8}
\affiliation{Department of Physics, University of Toronto, 60 St. George Street, Toronto, ON M5S 1A7}
\affiliation{David A. Dunlap Department of Astronomy, University of Toronto, 50 St. George Street, Toronto, ON M5S 3H4}

\author{Michael W. Coughlin}
\affiliation{Minnesota Institute for Astrophysics, University of Minnesota, Minneapolis, MN 55455, USA}
\affiliation{School of Physics and Astronomy, University of Minnesota, Minneapolis, MN 55455, USA}

\author{Michael Zevin}
\affiliation{The Adler Planetarium, 1300 South DuSable Lake Shore Drive, Chicago, IL, 60605, USA}
\affiliation{Center for Interdisciplinary Exploration and Research in Astrophysics (CIERA), Northwestern University, 1800 Sherman Ave., Evanston, IL, 60201, USA}
\affiliation{NSF-Simons AI Institute for the Sky (SkAI), 172 E. Chestnut St., Chicago, IL 60611, USA}

\author{Deep Chatterjee}
\affiliation{LIGO Laboratory, 185 Albany St, MIT, Cambridge, MA 02139, USA}

\author{Teagan A. Clarke}
\affiliation{School of Physics and Astronomy, Monash University, VIC 3800, Australia}
\affiliation{OzGrav: The ARC Centre of Excellence for Gravitational-wave Discovery, Clayton, VIC 3800, Australia}

\author{Storm Colloms}
\affiliation{Institute for Gravitational Research, University of Glasgow, Kelvin Building, University Avenue, Glasgow, G12 8QQ, Scotland}

\author{Utkarsh Mali}
\affiliation{Department of Physics, University of Toronto, 60 St. George Street, Toronto, ON M5S 1A7}
\affiliation{Canadian Institute for Theoretical Astrophysics, University of Toronto, 60 St. George Street, Toronto, ON M5S 3H8}

\author{Simona Miller}
\affiliation{Department of Physics, California Institute of Technology, Pasadena, California 91125, USA}
\affiliation{LIGO Laboratory, California Institute of Technology, Pasadena, California 91125, USA}

\author{Nathan Steinle}
\affiliation{Department of Physics and Astronomy, University of Manitoba, Winnipeg, MB R3T 2N2, CA}
\affiliation{Winnipeg Institute for Theoretical Physics, University of Manitoba, Winnipeg, MB R3T 2N2, CA}

\author{Pratyusava Baral} 
\affiliation{Leonard E. Parker Center for Gravitation, Cosmology, and Astrophysics, University of Wisconsin-Milwaukee, Milwaukee, WI 53201, USA}

\author{Amanda C. Baylor}    
\affiliation{Leonard E. Parker Center for Gravitation, Cosmology, and Astrophysics, University of Wisconsin-Milwaukee, Milwaukee, WI 53201, USA}

\author{Gareth {Cabourn Davies}} 
\affiliation{Institute of Cosmology and Gravitation, University of Portsmouth, Portsmouth, PO1 3FX, UK}

\author{Thomas Dent}      
\affiliation{IGFAE, University of Santiago de Compostela, 15872 Spain}

\author{Prathamesh Joshi} 
\affiliation{Department of Physics, The Pennsylvania State University, University Park, PA 16802, USA}
\affiliation{Institute for Gravitation and the Cosmos, The Pennsylvania State University, University Park, PA 16802, USA}
\affiliation{School of Physics, Georgia Institute of Technology, Atlanta, GA 30332, USA}

\author{Praveen Kumar}    
\affiliation{IGFAE, University of Santiago de Compostela, 15872 Spain}


\author{Cody Messick}     
\affiliation{Leonard E. Parker Center for Gravitation, Cosmology, and Astrophysics, University of Wisconsin-Milwaukee, Milwaukee, WI 53201, USA}

\author{Tanmaya Mishra}   
\affiliation{Department of Physics, University of Florida, PO Box 118440, Gainesville, FL 32611-8440, USA}


\author{Amazigh Ouzriat}  
\affiliation{Institut de Physique des 2 Infinis de Lyon (IP2I) - UMR 5822, Université de Lyon, Université Claude Bernard, CNRS, F-69622 Villeurbanne, France}

\author{Khun Sang Phukon} 
\affiliation{School of Physics and Astronomy and Institute for Gravitational Wave Astronomy, University of Birmingham, Edgbaston, Birmingham, B15 2TT, United Kingdom}

\author{Lorenzo Piccari} 
\affiliation{Dipartimento di Fisica, Università di Roma ''Sapienza'', Piazzale A. Moro 5, I-00185, Rome, Italy INFN Sezione di Roma, Piazzale A. Moro 5, I-00185, Rome, Italy}

\author{Marion Pillas} 
\affiliation{STAR Institute, Liege University, Allée du Six Août, 19C B-4000 Liege, Belgium}

\author{Max Trevor} 
\affiliation{Department of Physics, University of Maryland, College Park, Maryland 20742}

\author{Thomas A. Callister}
\affiliation{Kavli Institute for Cosmological Physics, The University of Chicago, 5640 S. Ellis Ave., Chicago, IL 60615, USA}

\author{Maya Fishbach}
\affiliation{Canadian Institute for Theoretical Astrophysics, University of Toronto, 60 St. George Street, Toronto, ON M5S 3H8}
\affiliation{Department of Physics, University of Toronto, 60 St. George Street, Toronto, ON M5S 1A7}
\affiliation{David A. Dunlap Department of Astronomy, University of Toronto, 50 St. George Street, Toronto, ON M5S 3H4}


\begin{abstract}
    We describe the effort to characterize gravitational-wave searches and detector sensitivity to different types of compact binary coalescences during the LIGO-Virgo-KAGRA Collaborations' fourth observing run.
    We discuss the design requirements and example use cases for this data product, constructed from \result{$> 4.33\times10^8$} injections during O4a alone.
    We also identify subtle effects with high confidence, like diurnal duty cycles within detectors.
    This paper accompanies a public data release of the curated injection set~\cite{GWTC-4-injections, GWTC-4-cumulative-injections}, and the appendixes give detailed examples of how to use the publicly available data.
\end{abstract}

\maketitle


\section{Introduction}
\label{sec:introduction}

\begin{figure}

    \begin{tikzpicture}[node distance=1.5cm]
        \tikzstyle{node} = [circle, text centered, draw=black, fill=white];
        \tikzstyle{recordednode} = [circle, text centered, draw=red, fill=red, fill opacity=0.1, text opacity=1.0];
        \tikzstyle{arrow} = [thick, ->, >=stealth, draw=black]
        \tikzstyle{recordedarrow} = [thick, ->, >=stealth, draw=red]
        %
        %
        \draw [fill=gray, opacity=0.2] (-2.00, -0.75) rectangle (+2.50, -11.00);
        \draw [draw=black] (-2.00, -0.75) rectangle (+2.50, -11.00);
        \node at (+1.60, -10.75) {$e=1...N_\mathrm{det}$};
        \node [rotate=90] at (-2.25, -5.35) {events};
        \draw [fill=gray, opacity=0.2] (-0.75, -3.35) rectangle (+2.25, -7.75);
        \draw [draw=black] (-0.75, -3.35) rectangle (+2.25, -7.75);
        \node at (+1.40, -7.50) {$i=1...N_\mathrm{ifo}$};
        \node [rotate=90] at (-1.00, -5.55) {detectors};
        %
        %
        \node (lambda) [node] {$\Lambda$};
        \node (theta) [recordednode, below left of=lambda, yshift=-0.5cm] {$\theta_e$};
        \draw [recordedarrow] (lambda) -- (theta);
        \node (t) [recordednode, below right of=lambda, yshift=-0.5cm] {$t_e$};
        \draw [arrow] (lambda) -- (t);
        \node (h) [node, below of=theta, xshift=-0.25cm, yshift=+0.15cm] {$h_{pe}$};
        \draw [arrow] (theta) -- (h);
        \draw [arrow] (t) -- (h);
        \node (antenna) [node, below right of=h, xshift=+0.25cm] {$F_{ip}$};
        \draw [arrow] (theta) -- (antenna);
        \draw [arrow] (t) -- (antenna);
        \node (strain) [node, below of=antenna] {$h_{ie}$};
        \draw [arrow] (h) -- (strain);
        \draw [arrow] (antenna) -- (strain);
        \node (n) [node, right of=strain, xshift=-0.40cm] {$n_{ie}$};
        \draw [arrow] (t) -- (n);
        \node (d) [node, below of=strain] {$d_{ie}$};
        \draw [arrow] (strain) -- (d);
        \draw [arrow] (n) -- (d);
        \node (F) [recordednode, below of=d, yshift=-0.25cm] {$\mathcal{F}_e$};
        \draw [arrow] (d) -- (F);
        \node (vartheta) [recordednode, left of=F, xshift=+0.25cm] {$\vartheta_e$};
        \draw [arrow] (d) -- (vartheta);
        \node (D) [node, below of=F] {$\mathbb{D}_e$};
        \draw [arrow] (F) -- (D);
        %
        %
        \node [right of=lambda, xshift=+2.2cm] {population};
        \node [right of=t, xshift=+2.0cm, yshift=+0.00cm] {single-event parameters};
        \node [right of=h, xshift=+3.1cm] {signal};
        \node [right of=antenna, xshift=+2.6cm, yshift=+0.00cm] {detector response};
        \node [right of=n, xshift=+1.4cm, yshift=+0.17cm] {projected strain};
        \node [right of=n, xshift=+1.5cm, yshift=-0.17cm] {\& detector noise};
        \node [right of=d, xshift=+1.8cm] {data};
        \node [right of=F, xshift=+2.5cm, yshift=+0.17cm] {search statistics};
        \node [right of=F, xshift=+3.1cm, yshift=-0.17cm] {\& signal reconstruction};
        \node [right of=D, xshift=+2.1cm] {detection};
    \end{tikzpicture}

    \caption{
        Directed acyclic graph (DAG) showing the conditional (in)dependencies between variables involved in the construction of a catalog~\cite{Essick:2023upv}.
        Circles (nodes) represent individual variates, and arrows (directed edges) denote conditional dependencies.
        Shaded rectangles (plates) represent independently and identically distributed variables (i.e., $N_\mathrm{det}$ events and $N_\mathrm{ifo}\times N_\mathrm{det}$ data vectors, one in each IFO for each event).
        Red nodes and edges represent values that are recorded within the public data products~\cite{GWTC-4-cumulative-injections, GWTC-4-injections}.
        Importantly, we record $p(\theta_e|\Lambda_\mathrm{inj})$ and $\theta_e$ for each event so that users can easily reweight samples to follow other populations within Monte Carlo sums; see Eq.~\ref{eq:Phat(D|Lambda)} and Appendix~\ref{sec:using the data product}.
    }
    \label{fig:dag}
\end{figure}
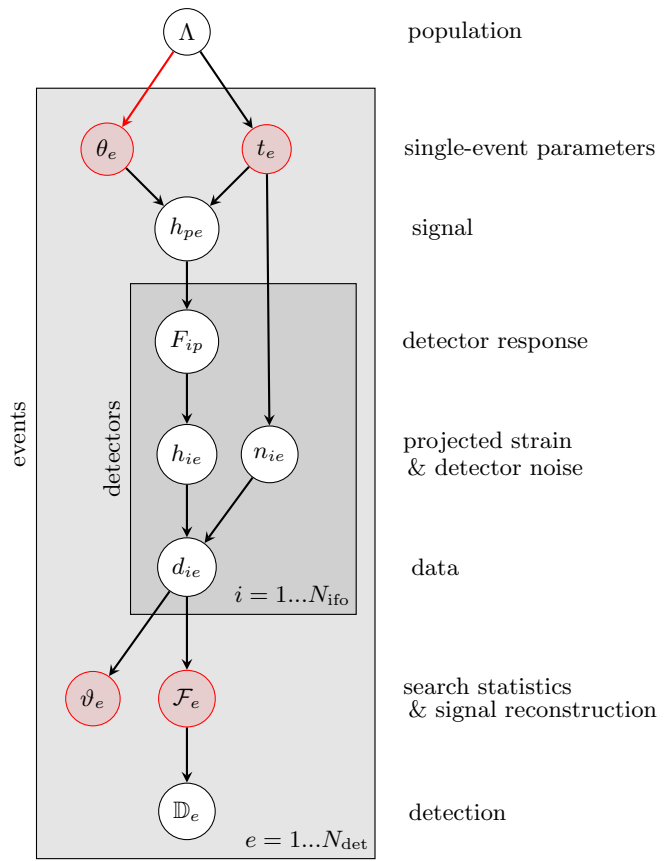

Ground-based gravitational-wave (GW) interferometers (IFOs), like advanced LIGO~\cite{LIGO}, advanced Virgo~\cite{Virgo}, and KAGRA~\cite{KAGRA}, have reached astonishing precision, yielding catalogs of several hundred confidently detected compact binary coalescences (CBCs) \cite{GWTC-1, GWTC-2, GWTC-2.1, GWTC-3} since the first direct detection in 2015 \cite{GW150914}.
In order to maximize the scientific return on these catalogs, we require a detailed understanding of the detection process.
That is, how astrophysical signals are recorded in the interferometers and how data is selected by GW searches.
We follow the notation introduced in \citet{Essick:2023upv}, which describes this process as a hierarchical model connecting the astrophysical population to the data produced in detectors and detection thresholds (Fig.~\ref{fig:dag}).
This is possible, in no small part, because of the work of many scientists over many years to characterize not only the response of the interferometers and their calibration but also theoretical modeling of the expected signal from CBCs with high precision.
Because of the decades of investment from both the experimental/instrumentation and theoretical communities, it is possible to build an accurate and precise forward model of the entire process of generating a GW catalog.

This paper describes the effort to precisely characterize that process and produce community data products for the LIGO-Virgo-KAGRA (LVK) Collaborations' fourth observing run (O4; \result{May 2023 -- November 2025}).
Specifically, this paper accompanies a data release of sensitivity estimates~\cite{GWTC-4-cumulative-injections, GWTC-4-injections}, both cumulative and for individual observing runs, and describes how these can be used to compute quantities of interest such as the detection probability (or efficiency) and the sensitive time-volume.
These data products can also be used to train classifiers based on the signal reconstructions and detection statistics reported by searches in low latency through straightforward density estimation~\cite{Chatterjee:2019avs, Chaudhary:2023vec}.

The LVK has previously released similar data products~\cite{O3a-injections, GWTC-3-injections, GWTC-3-cumulative-injections}, and at least one external group likewise processed injections spanning the third observing run (O3)~\cite{Mehta:2025jiq}.
For O4, we made several improvements to the injection campaign's design and expanded the data included within the final product.
Specifically, we
\begin{itemize}
    \item updated the injected distribution to broaden the coverage of parameter space and remove ``holes'' that were present in previous campaigns (e.g., high-spin and low-mass signals were not included in Refs.~\cite{GWTC-3-injections, GWTC-3-cumulative-injections}; see also Appendix~\ref{sec:low-dim sampling}),
    \item used injected distributions that resemble the astrophysical distributions inferred previously while maintaining relatively broad tails to support possible outliers,
    \item provided sensitivity results from all searches at all masses,
    \item provided additional search statistics for each detected injections, including parameters describing the reconstructed signal (e.g., best-match masses and spins for template-based searches), and
    \item included significantly more semianalytic injections \cite{Essick:2023toz} for O1 and O2 (with distributions that match O4 in everything but the maximum redshift).
\end{itemize}
The results of the injection campaign for O4a (\result{May 2023 -- January 2024}) are available publicly within Zenodo~\cite{GWTC-4-cumulative-injections, GWTC-4-injections}.
Additional results will be released for O4b (\result{March 2024 -- January 2025}) and O4c (\result{January 2025 -- November 2025}) at the same time as those catalogs.

This paper is structured as follows.
Sec.~\ref{sec:model for catalog generation} describes our model for how catalogs are generated.
Sec.~\ref{sec:injection campaign scope} describes the scope of the injection campaign during O4, and Sec.~\ref{sec:injection distribution} describes the injected distribution and how it was chosen.
It also compares the injected and recovered distributions, including identifying well-known correlations between masses, distances, and the orbital inclination induced by conditioning on detection along with more subtle effects like diurnal cycles within GW IFOs and the associated selection on the right ascension of detected sources.
Sec.~\ref{sec:injection generation workflow} briefly describes the workflow used to generate the injections and the size of the dataset; during O4a, we generated \result{433,157,717} waveforms, performed \result{3,974,400} injections, and recovered \result{476,415} signals with a realistic false alarm rate (FAR) threshold of \result{1/year}.
Injection sets of similar size are being constructed for O4b and O4c.
Finally, we conclude in Sec.~\ref{sec:discussion}.
The Appendixes serve as a one-stop-shop with numerical recipes for how to use the public data products.


\section{Hierarchical Model for Catalog Generation}
\label{sec:model for catalog generation}

We follow the discussion in \citet{Essick:2023upv} and model the astrophysical process through which CBCs form and merge as an inhomogeneous Poisson process with differential rate density
\begin{equation}\label{eq:dN/dtheta}
    \frac{dN}{d\theta dt} = \mathcal{K} p(\theta, t|\Lambda)
\end{equation}
In this notation, parameters of individual CBCs (masses, spins, redshift, etc) are denoted by $\theta$, the detector-frame time at which an event passes through the center of the Earth is denoted by $t$,\footnote{This could be included as one of the individual parameters of the signal ($\theta$), but we call it out separately for clarity.} the shape of the population is described by hyperparameters $\Lambda$, and the overall number of astrophysical binaries is given by $\mathcal{K}$.
Given a fixed number of events, individual CBCs are independently and identically drawn from $p(\theta, t|\Lambda)$.
Each CBC will produce a GW strain in polarization ($p \in \{+, \times\}$: $h_p(\theta)$) to which each GW interferometer ($i \in \{\mathrm{LHO}, \mathrm{LLO}, \mathrm{Virgo}, \ldots$) will respond according to the detector response $F_{ip}(\theta)$, which typically is modeled as a linear transfer function that depends on the relative orientation of the source and the detector (see, e.g., \citet{Essick:2017wyl}).
Data in each interferometer is assumed to be a linear combination of the projected strain and additive noise: $d_i = n_i + F_{ip} h_p$.
Finally, searches process this data and compute detection statistics ($\mathcal{F}$) like the signal-to-noise ratio ($\rho$) and FAR as well as signal reconstructions parametrized by $\vartheta$ (e.g., best-match template masses and spins).
Whether an event is detected ($\mathbb{D}$) is then determined by whether the search statistic exceeds a threshold ($\mathcal{F} > \mathcal{F}_\mathrm{thr}$).
Fig.~\ref{fig:dag} shows the conditional (in)dependencies within this process.

Taken as a whole, the model in Fig.~\ref{fig:dag} corresponds to a joint distribution over the true parameters of individual events ($\theta_e, t_e$), the noise in each detector at the time of the event ($n_{ie}$), the data recorded by each detector ($d_{ie}$), the statistics ($\mathcal{F}_e$) and reconstructions ($\vartheta_e$) produced by searches, and whether events were detected ($\mathbb{D}_e$), all conditioned on the astrophysical population ($\Lambda$).
For a catalog containing $N_\mathrm{det}$ events (indexed by $e$) observed with $N_\mathrm{ifo}$ interferometers (indexed by $i$), this is
\begin{widetext}
\begin{multline}\label{eq:joint}
    p(\left\{\theta_e, t_e, \{n_{ie}, d_{ie}\}_{i=1...N_\mathrm{ifo}}, \mathcal{F}_e, \vartheta_e, \mathbb{D}_e\right\}_{e=1...N_\mathrm{det}}, \mathcal{K}|N_\mathrm{det}, N_\mathrm{ifo}, \Lambda) \\
        = \mathcal{K}^{N_\mathrm{det}} e^{-\mathcal{K}P(\mathbb{D}|\Lambda)} \prod\limits_e^{N_\mathrm{det}} \left[ p(\theta_e, t_e|\Lambda) \left[ \prod\limits_i^{N_\mathrm{ifo}} p(n_{ie}|t_e) p(d_{ie}|n_{ie},\theta_e,t_e) \right] p(\mathcal{F}_e, \vartheta_e|\left\{d_{ie}\right\}_{i=1...N_\mathrm{ifo}}) P(\mathbb{D}_e|\mathcal{F}_e) \right]
\end{multline}
\end{widetext}
where we have again called out the time of the event ($t$) separately from the rest of the event parameters ($\theta$) for clarity later on.
Eq.~\ref{eq:joint} also introduces the detection probability
\begin{widetext}
\begin{align}
    P(\mathbb{D}|\Lambda)
        & = \int d\theta dt\, p(\theta, t|\Lambda) \int \prod\limits_i^{N_\mathrm{ifo}} \mathcal{D}n_i \mathcal{D} d_i\, p(n_i|t) p(d_i|n_i,\theta,t) \int d\mathcal{F} p(\mathcal{F}|\{d_i\}) P(\mathbb{D}|\mathcal{F}) \label{eq:P(D|Lambda)}
\end{align}
\end{widetext}
which quantifies the probability that an individual event drawn from the population would be detected.

We additionally assume that the data is a deterministic function of the noise and signal and employ models for the waveform $h_p(\theta)$ (see Appendix~\ref{sec:waveform choices}) and noise processes $p(n_{i}|t)$ (see Appendix~\ref{sec:psd variability}).
Noise is often modeled as stationary, Gaussian, and independent between interferometers.
Similarly, it is common to model the search statistics as deterministic functions of the data ($\mathcal{F}_e = \mathcal{F}(\left\{d_{ie}\right\})$) and detection as a deterministic function of the search statistics with a simple threshold ($\mathbb{D}_e = \Theta(\mathcal{F}_e > \mathcal{F}_\mathrm{thr})$, where $\Theta(\cdot)$ is an indicator function: one when the argument is true and zero otherwise).
However, these assumptions are not strictly necessary.
Again, see \citet{Essick:2023upv} for more details.

With this model in hand, we can construct inferences for the astrophysical population conditioned on observed data: $p(\Lambda|\left\{d_{ie},\mathbb{D}_{e}\right\}, N_\mathrm{det}, N_\mathrm{ifo})$.
A crucial aspect of this is estimating Eq.~\ref{eq:P(D|Lambda)}, which encapsulates the sensitivity of searches to different populations (see, e.g., \citet{GWTC-3-RnP}).
Specifically, we wish to efficiently compute the high-dimensional integral in Eq.~\ref{eq:P(D|Lambda)}, which is commonly abbreviated as
\begin{equation}\label{eq:P(D|Lambda) short}
    P(\mathbb{D}|\Lambda) = \int d\theta\, p(\theta|\Lambda) P(\mathbb{D}|\theta,\Lambda)
\end{equation}
where
\begin{widetext}
\begin{align}\label{eq:P(D|theta,Lambda)}
    P(\mathbb{D} |\theta,\Lambda) & = \int dt p(t|\Lambda) \int \prod\limits_i^{N_\mathrm{ifo}}\mathcal{D}n_i \mathcal{D}d_i p(n_i|t) p(d_i|n_i,\theta,t) \int d\mathcal{F} p(\mathcal{F}|\{d_i\}) P(\mathbb{D}|\mathcal{F})
\end{align}
\end{widetext}
is the probability of detecting a particular signal ($\theta$) averaged over the noise throughout the experiment, and we have assumed $p(\theta,t|\Lambda) = p(\theta|\Lambda) p(t|\Lambda)$.
It is also common to assume that $p(t|\Lambda)$ is always uniform and to drop the dependence on $\Lambda$, in which case Eq.~\ref{eq:P(D|theta,Lambda)} is written as simply $P(\mathbb{D}|\theta)$.
Importantly, as \citet{Essick:2023upv} describes, this expression \emph{must} be interpreted as an implicit integral over noise realizations and not as a statement that detection depends directly on the true signal parameters.

What's more, we may not know exactly how some variables are distributed.
For example, while we can often model noise as stationary and Gaussian over short periods of time, we do not know how it will vary over the entire length of the run.
As such, we adopt a Monte Carlo approximation for these integrals and generate many draws from a reference distribution:
\begin{align}
    \theta_e \sim p(\theta|\Lambda_\mathrm{inj}) \\
    t_e \sim p(t|\Lambda_\mathrm{inj})
\end{align}
These simulated signals are injected throughout the run ($p(t|\Lambda_\mathrm{inj})$ is uniform in detector-frame wall-time) in order to sample over different noise realizations generated by the real noise processes and detector duty cycles~\cite{Chen:2016luc}.
Searches then process the corresponding data, reporting detection statistics and signal reconstructions.
These can be compared against different detection thresholds to determine which events would have been included within a catalog.

In this way, we can approximate the integral in Eq.~\ref{eq:P(D|Lambda) short} for an arbitrary population with a Monte Carlo (importance sampling) sum
\begin{align}
    P(\mathbb{D}|\Lambda)
        & = \int d\theta\, p(\theta|\Lambda) P(\mathbb{D}|\theta) \nonumber \\
        & = \int d\theta\, p(\theta|\Lambda_\mathrm{inj}) \left[\frac{p(\theta|\Lambda)}{p(\theta|\Lambda_\mathrm{inj})} P(\mathbb{D}|\theta)\right] \nonumber \\
        & \approx \hat{P}(\mathbb{D}|\Lambda) \equiv \frac{1}{N_\mathrm{inj}} \sum\limits_{e}^{N_\mathrm{inj}} \left[ \frac{p(\theta_e|\Lambda)}{p(\theta_e|\Lambda_\mathrm{inj})} \mathbb{D}_e \right] \label{eq:Phat(det|Lambda) long}
\end{align}
where, again, $\theta_e \sim p(\theta|\Lambda_\mathrm{inj})$ and $\mathbb{D}_e$ records whether an individual injection passes the detection threshold ($\mathcal{F}_e > \mathcal{F}_\mathrm{thr}$).
Because only the $N_\mathrm{fnd} \leq N_\mathrm{inj}$ injections that pass the detection threshold contribute to the sum, this expression is often shortened to
\begin{equation}\label{eq:Phat(D|Lambda)}
    \hat{P}(\mathbb{D}|\Lambda) = \frac{1}{N_\mathrm{inj}} \sum\limits_{e}^{N_\mathrm{fnd}} \left[ \frac{p(\theta_e|\Lambda)}{p(\theta_e|\Lambda_\mathrm{inj})} \right]
\end{equation}
The public data products~\cite{GWTC-4-cumulative-injections, GWTC-4-injections} record both $\theta_e$ and $p(\theta_e|\Lambda_\mathrm{inj})$ for each injection along with $N_\mathrm{inj}$ so that users can easily construct sums like Eq.~\ref{eq:Phat(D|Lambda)}.
See Appendix~\ref{sec:using the data product} for more discussion and other Monte Carlo techniques.

As is evident from Eq.~\ref{eq:Phat(det|Lambda) long}, this approximation is only valid if the support of $p(\theta|\Lambda)$ is entirely contained within the support of $p(\theta|\Lambda_\mathrm{inj})$.
Otherwise, we would effectively multiply and divide by zero within some region of $\theta$-space, and the Monte Carlo sum would suffer from a ``silent truncation error.''
The sum in Eq.~\ref{eq:Phat(D|Lambda)} would still be well-defined, but it would no longer be an accurate approximation to Eq.~\ref{eq:P(D|Lambda) short}.

With this caveat in mind, we can estimate the detection probability for many different populations with a single injection set.
This means that searches do not have to process separate injection sets for each population, greatly reducing the computational cost of the hierarchical Bayesian inference.
However, the estimator $\hat{P}(\mathbb{D}|\Lambda)$ can be imprecise if $p(\theta|\Lambda)$ differs significantly from $p(\theta|\Lambda_\mathrm{inj})$ (i.e., when there are very few effective samples; see, e.g., \citet{Farr:2019rap} and~\citet{Essick:2022ojx}).
Precision can always be improved by increasing the injection size, but this can require significant additional computational cost.

It is therefore crucial to design injection campaigns so that they can be used to estimate $P(\mathbb{D}|\Lambda)$ precisely enough to support population inferences with large catalogs while at the same time maintaining a small enough computational cost to be tractable for searches.


\subsection{Approximations and Other Approaches}
\label{sec:approximations and other approaches}

Before progressing to the details of the O4 injection campaign, it is worth mentioning a few other approximations that have been explored in the literature.
Several authors have \textit{de facto} proposed alternative methods to estimate $P(\mathbb{D}|\theta)$ assuming $p(t|\Lambda)$ is uniform.

Some authors have fit flexible models to $P(\mathbb{D}|\theta)$ with neural networks \cite{Callister:2024qyq} or analytic fitting functions \cite{Lorenzo-Medina:2024opt} using existing injection sets as training data.
Similarly, others model $p(\theta|\mathbb{D},\Lambda_\mathrm{inj}) \propto P(\mathbb{D}|\theta) p(\theta|\Lambda_\mathrm{inj})$, such as \citet{Talbot:2020oeu} who use a Gaussian mixture model in a transformed coordinate system.
Both approaches are predicated upon the existence of large injection campaigns for their training data.
Additionally, they make few assumptions about the physical processes underlying detection, and therefore may not generalize well to other observing runs, detector sensitivities, or data sets that were not included in their training data.

Semianalytic sensitivity estimates~\cite{Finn:1992xs, Essick:2023toz}, on the other hand, typically assume the noise processes are independent in each detector and described by stationary Gaussian noise over short periods, which is often a reasonable approximation.
This yields a tractable model for $p(n|t)$.
It is also common to assume the noise distributions are described by a single reference power spectral density (PSD) for long stretches of time (e.g., months or years), which may be less justified.
Additionally, various flavors of semianalytic models approximate detection based on the signal-to-noise ratio (either optimal or observed) with a simple threshold~\cite{Essick:2023toz}.\footnote{\citet{Essick:2023toz} and \citet{Essick:2023upv} show that the most accurate and consistent semianalytic selection estimate is based on the the observed signal-to-noise ratio ($\rho_\mathrm{obs}$), also sometimes called the matched-filter signal-to-noise ratio.}
Unlike other approaches, this leaves a single free parameter: the threshold $\rho_\mathrm{thr}$.
While such approaches have been shown to closely match real injection campaigns and are expected to generalize well~\cite{Essick:2023toz}, they nevertheless still require real injection campaigns to determine an appropriate $\rho_\mathrm{thr}$ that matches different thresholds on real search statistics.

Regardless of whether one uses real injections or one of these approximations (all of which rely in some way on real injections), we again remind the reader that selection effects must be taken into account directly within the inference.
One cannot ``fit the detected distribution'' and then ``divide by the detection probability'' \textit{post hoc} within an inference framework that is consistent with physical detection processes~\cite{Essick:2023upv}.
As such, injection campaigns are an unavoidable and invaluable part of population inference from observed catalogs.


\section{Campaign Scope}
\label{sec:injection campaign scope}

Given the probabilistic model of catalog construction in Sec.~\ref{sec:model for catalog generation} and the Monte Carlo approximation for the associated high-dimensional integrals, we still need to determine the number of injections required in order to make those sums precise estimates.
Typically, this depends on the catalog size (larger catalogs require more precision and therefore more injections~\cite{Farr:2019rap, Essick:2022ojx}), so we first discuss the expected size of the catalog at the end of O4.


\subsection{The Expected Catalog Size after O4}
\label{sec:the expected catalog size after o4}

\begin{figure*}
    \includegraphics[width=1.0\textwidth]{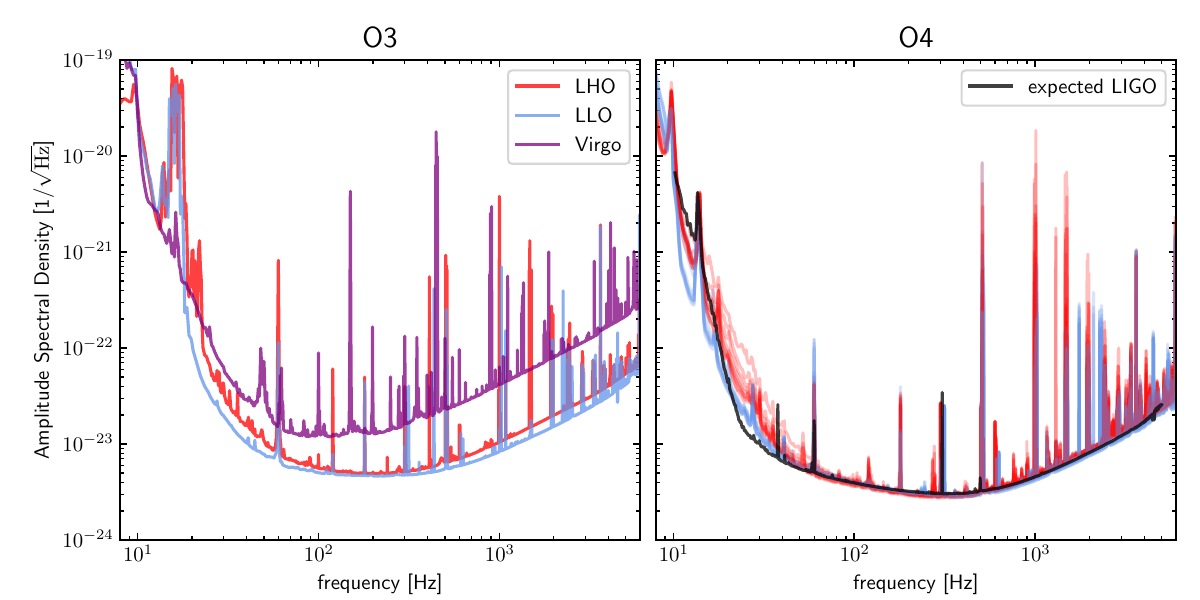}
    \caption{
        Reference power spectral densities (PSDs) used when constructing the O4 injection set and benchmarking the number of expected events.
        (\emph{left}) Observed PSDs during O3 and (\emph{right}) expected and observed PSDs during O4a.
        There is a separate trace for each LIGO's PSD during each month of O4a (\result{May 2023 -- January 2024}) as the injection sets were created on a monthly basis.
        See Appendix~\ref{sec:psd variability} for more discussion about the impact of PSD variability.
    }
    \label{fig:psd}
\end{figure*}

Given the inhomogeneous Poisson rate density (Eq.~\ref{eq:dN/dtheta}), the expected number of detected events within an observing run is
\begin{align}
    \mathrm{E}[N_\mathrm{det}]
        & = \mathcal{K} \int d\theta p(\theta|\Lambda) P(\mathbb{D}|\theta) \\
        & = \mathcal{R} T P(\mathbb{D}|\Lambda)
\end{align}
where we have reexpressed $\mathcal{K}$ as an overall rate of astrophysical events per detector-frame time ($\mathcal{R}$) and the observing run's duration ($T$).
By noting that neither $\mathcal{R}$ nor $p(\theta|\Lambda)$ will change between runs, we can predict the number of detections in a future run ($A$) based on a past run ($B$) via\footnote{Obviously, this can also be expressed as the expected rate of detections per detector-frame time rather than the number of detections.}
\begin{equation}
    \frac{\mathrm{E}[N_A|\Lambda]}{\mathrm{E}[N_B|\Lambda]} = \left(\frac{T_A}{T_B}\right) \left(\frac{P_A(\mathbb{D}|\Lambda)}{P_B(\mathbb{D}|\Lambda)}\right)
\end{equation}
A more careful estimate would also marginalize over uncertainty in the astrophysical distribution.
However, for expediency, we pick as a representative the injected distribution described in Sec.~\ref{sec:distribution}.\footnote{It is also possible to consider subpopulations of particular types of sources, such as binary neutron star (BNS) mergers.}

Before the beginning of O4, we estimated the detection probabilities assuming representative PSDs measured during O3 and predicted for O4 (\citet{GWTC-4-cumulative-injections} and~\citet{O4-projected-PSD}, respectively); see Fig.~\ref{fig:psd}.
We estimate $P(\mathbb{D}|\Lambda)$ semianalytically with a detection threshold of \result{10} on the observed phase-maximized network signal-to-noise ratio~\cite{Essick:2023toz}.
Additionally, we take O3's duration to be \result{$T_\mathrm{O3} = 12\,\mathrm{months}$} (accounting for commissioning breaks, duty cycles, etc) and O4's expected duration to be \result{$T_\mathrm{O4} = 25\, \mathrm{months}$} (assuming similar detector duty cycles).

\begin{figure}
    \includegraphics[width=1.0\columnwidth]{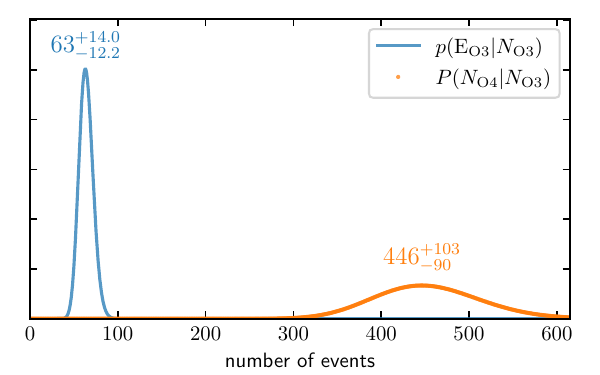}
    \caption{
        The distribution of the expected number of detections in O3 ($E_\mathrm{O3}$) given the observed number of detection in O3 (\externalresult{$N_\mathrm{O3}=63$ with FAR $<$ 1/year}) and the distribution of the observed number of detections in 25 months of O4 ($N_\mathrm{O4}$) given $N_\mathrm{O3}$ based on semianalytic estimates for $P(\mathbb{D}|\Lambda)$~\cite{Essick:2023toz}. 
        Annotations denote the maximum-likelihood and 90\% highest-probability-density credible regions.
        We denote the distribution over $E_\mathrm{O3}$ with a line because it is continous, and we denote the distribution over $N_\mathrm{O4}$ with dots because it is discrete.
    }
    \label{fig:expected numbers}
\end{figure}

Finally, we estimate the expected number of detections during O3 ($\mathrm{E}_\mathrm{O3}$) from the observed number of detections (\externalresult{$N_\mathrm{O3} = 63$ with FAR $<$ 1/year} \cite{GWTC-3-RnP}) assuming Poisson counting uncertainty
\begin{equation}\label{eq:p(EO3|NO3)}
    p(\mathrm{E}_\mathrm{O3}|N_\mathrm{O3}) = \frac{\mathrm{E}_\mathrm{O3}^{N_\mathrm{O3}}}{\Gamma(N_\mathrm{O3}+1)} e^{-\mathrm{E}_\mathrm{O3}}
\end{equation}
and
\begin{equation}
    P(N_\mathrm{O4}|\mathrm{E}_\mathrm{O4}) = \frac{\mathrm{E}_\mathrm{O4}^{N_\mathrm{O4}}}{N_\mathrm{O4}!} e^{-\mathrm{E}_\mathrm{O4}}
\end{equation}
so that
\begin{align}
    P(N_\mathrm{O4}|N_\mathrm{O3})
        & = \int d\mathrm{E}_\mathrm{O3} p(\mathrm{E}_\mathrm{O3}|N_\mathrm{O3}) P(N_\mathrm{O4}|\mathrm{E}_\mathrm{O4}=C\mathrm{E}_\mathrm{O3}) \nonumber \\
        & = \frac{C^{N_\mathrm{O4}}}{(1+C)^{N_\mathrm{O3} + N_\mathrm{O4} + 1}} \left( \begin{matrix} N_\mathrm{O3} + N_\mathrm{O4} \\ N_\mathrm{O3} \end{matrix} \right) \label{eq:P(NO4|NO3)}
\end{align}
where \result{
\begin{align}\label{eq:C}
    C
        & = \left(\frac{T_\mathrm{O4}}{T_\mathrm{O3}}\right) \left(\frac{P_\mathrm{O4}(\mathbb{D}|\Lambda)}{P_\mathrm{O3}(\mathbb{D}|\Lambda)}\right) \nonumber \\
        & \approx \left(\frac{25}{12}\right) \left(\frac{3.51\times10^{-3}}{1.03\times10^{-3}}\right) \approx 7.1
\end{align}
}
Fig.~\ref{fig:expected numbers} shows the distributions of the numbers of events within each run.
We expect to observe \result{$N_\mathrm{O4} = 446^{+103}_{-90}$ events} during \result{25 months} of O4 (maximum likelihood and 90\% highest-probability-density interval).
However, it is worth noting that the detectors did not quite reach their anticipated sensitivities during O4 (Fig.~\ref{fig:psd}), so this is an optimistic estimate.


\subsection{Requirements for the Number of Injections During O4}
\label{sec:requirements for the number of injections during o4}

Several arguments~\cite{Farr:2019rap, Essick:2022ojx} suggest that stable hierarchical inference requires $N_\mathrm{eff} \gtrsim 4 N_\mathrm{det}$, where $N_\mathrm{eff}$ is the effective number of samples defined as the ratio of the expected value of $\hat{P}(\mathbb{D}|\Lambda)$ squared to its variance under Monte Carlo uncertainty
\begin{equation}
    N_\mathrm{eff} = \frac{\mathrm{E}[\hat{P}(\mathbb{D}|\Lambda)]^2}{\mathrm{V}[\hat{P}(\mathbb{D}|\Lambda)]}
\end{equation}
At the end of O4, we expect to have at most \result{$N_\mathrm{det} \leq 612 = 63 + 549$ events} (at 90\% credibility), meaning we will require \result{$N_\mathrm{eff} \gtrsim 2,448$} as a bare minimum.
A safer estimate would be \result{$N_\mathrm{eff} \gtrsim 10^4$}.

High-dimensional population models can often produce effective numbers of samples that are only a few percent of the number of detected injections ($N_\mathrm{eff} \sim 10^{-2} N_\mathrm{fnd}$).
This means that we will likely need \result{$N_\mathrm{fnd} \gtrsim 10^6$} to obtain \result{$N_\mathrm{eff} \gtrsim 10^4$}.

As a rough approximation, we use the number of detectable injections as those with an optimal network signal-to-noise ratio \result{$\rho_\mathrm{opt,net} \geq 10$}.
If injections are only kept above a hopeless signal-to-noise cut $\rho_\mathrm{cut}$ (see~\citet{Essick:2023toz}) and the fraction of those injections above a threshold $\rho_\mathrm{thr} > \rho_\mathrm{cut}$ scales approximately as
\begin{equation}
    P(\rho > \rho_\mathrm{thr}) \sim \left(\frac{\rho_\mathrm{cut}}{\rho_\mathrm{thr}}\right)^3
\end{equation}
(strictly true only for a Euclidean universe or nearby sources, but a good approximation with current sensitivity), then this means we need to perform \result{$8\times10^6$} injections with \result{$\rho_\mathrm{opt,net} > \rho_\mathrm{cut} = 5$} in order to obtain \result{$N_\mathrm{fnd} \sim 10^6$} and consequently \result{$N_\mathrm{eff} \gtrsim 10^4$}.

Taking O3 as an example, we can expect at least one IFO to be observing for \result{80\%} of the time (\citet{GWTC-3} reports that at least one IFO was observing \externalresult{96.8\%} of the time, and at least two were observing \externalresult{81.8\%} of the time).
Therefore, using a conservative estimate for the amount of livetime during which searches may be sensitive (\result{$52.7\times10^6$ sec}), 
this means that injections that are actually performed ($\rho_\mathrm{opt,net} \geq \rho_\mathrm{cut}$) must be spaced by
\result{
\begin{equation}
    \Delta t \approx \frac{52.7\times10^6\,\mathrm{sec}}{8\times10^6\, \mathrm{events}} = 6.6\,\mathrm{sec}
\end{equation}
}

Note that we have included the events from O1, O2, and O3 in the total catalog size after O4 and therefore the projection for the required $N_\mathrm{eff}$ at the end of O4.
In reality, we already have injections from O1, O2, and O3.
If we only include the expected number of events from O4 but otherwise adopt the same factors of safety, this will reduce the number of injections needed by \result{$\lesssim 11\%$} (\result{$7.1\times10^6$} injections with \result{$\Delta t = 7.4\, \mathrm{sec}$}).

This is a relatively high rate of injections.
Even though the majority of events will have $\rho_\mathrm{opt,net}$ near $\rho_\mathrm{cut}$ and we can distinguish between the coalescence times of neighboring events much more precisely than $1\,\mathrm{sec}$, concerns about biasing PSD estimation within searches limited the allowed injection spacing to \result{$\Delta t \geq 24\,\mathrm{sec}$}.
As such, we constructed four separate injection streams that are processed independently throughout O4.
Each subset had \result{$24\ \mathrm{sec}$} spacing between injections, and the subsets are staggered so that the cumulative injection set has an effective spacing of \result{$6\ \mathrm{sec}$ ($\sim 8.8\times10^6$ injections with $\rho_\mathrm{opt,net} > \rho_\mathrm{cut}$)}.
See Sec.~\ref{sec:injection generation workflow} for more details.


\section{Injected Distribution and Distribution Design}
\label{sec:injection distribution}

We now describe the injected distribution developed for O4 along with the requirements that drove design choices.
Readers who are only interested in the final injected distribution can proceed directly to Sec.~\ref{sec:distribution}.


\subsection{Requirements for Parameter Coverage}
\label{sec:requirements for parameter coverage}

We begin by considering the range of masses that our injected distribution needs to cover in order to support inference of the Hubble constant (and other cosmological parameters; see Appendix~\ref{sec:using the data product}).
We do this by identifying a range of detector-frame masses with horizons large enough that we might detect at least one event within O4. 
We can then compute the source-frame mass range required to span that detector-frame mass range assuming the single reference cosmology used when generating injections.

To begin, we note that we approximate the sensitive volume for events with small horizon distances as if the universe was Euclidean
\begin{equation}
    V \approx \frac{4}{3}\pi D_L^3
\end{equation}
where $D_L$ is the horizon distance.
Given this, we estimate the expected number of detections given an astrophysical rate ($\mathcal{R}$) and an observation time ($T$) via
\begin{equation}
    \mathrm{E}[N] \approx \mathcal{R} T \frac{4}{3}\pi D_L^3
\end{equation}
Now, if we want to have support wherever there is a chance of detecting an event during O4, this can be accomplished by guaranteeing support wherever $\mathrm{E}[N] > \mathrm{E}_\mathrm{min} \sim 0.1$.
This means we should require
\begin{equation}
    D_L \gtrsim \left( \frac{3 \mathrm{E}_\mathrm{min}}{4 \pi \mathcal{R} T} \right)^{1/3}
\end{equation}
Taking a conservative estimate for the rate \result{$\mathcal{R} \sim 10^4 \, \mathrm{Gpc}^{-3} \mathrm{yr}^{-1}$} and \result{25 months of O4}, we obtain \result{$D_L \sim 10\,\mathrm{Mpc}$}.

\begin{figure}
    \includegraphics[width=1.0\columnwidth]{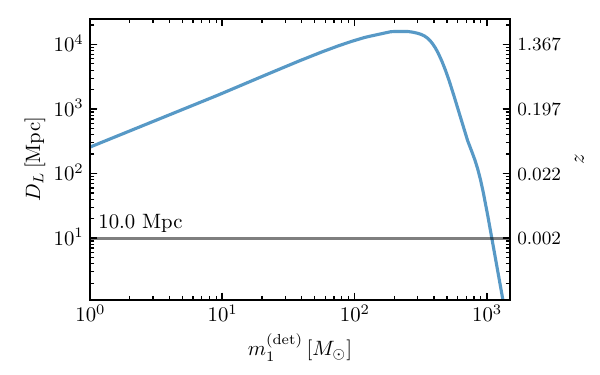}
    \caption{
        Horizon (\emph{left axis}) distance and (\emph{right axis}) redshift as a function of detector-frame primary mass for equal-mass, maximally spinning/aligned-spin binaries using the \result{expected LIGO PSD for O4} \cite{O4-projected-PSD} with \result{$\rho_\mathrm{opt,net} = 10$}.
    }
    \label{fig:horizons}
\end{figure}

Fig.~\ref{fig:horizons} shows the range of detector-frame primary masses that have horizon distances \result{$> 10\,\mathrm{Mpc}$} assuming equal-mass, optimally oriented binaries and a threshold of \result{$\rho_\mathrm{opt} > 10$} with respect to the predicted LIGO O4 PSD (Fig.~\ref{fig:psd}).
That is, we find that the lower limit of our mass range will be set by astrophysical priors (i.e., we do not believe there will be a significant number of signals below $\sim 1\,M_\odot$) while the upper limit corresponds to a detector-frame component mass of \result{$\sim 10^3\,M_\odot$}.
Because source-frame masses will always be smaller than detector frame masses, we conclude that distributions with support for source-frame component masses between \result{1--$10^3\, M_\odot$} should suffice for O4.

Given this mass range, we additional covered the entire domain of spin magnitudes allowed by cosmic censorship (\result{$0 \leq s \leq 1$}), all spin orientations, and the maximum redshift to be far enough away that it does not truncate the detected distribution.

We do not inject signals with tides (i.e., we only inject vacuum solutions at all masses), and we only inject quasi-circular signals (zero eccentricity).


\subsection{Additional Requirements}
\label{sec:additional requirements}

We also consider a few qualitative requests intended to make the data product easier to use (see Appendix~\ref{sec:using the data product}).
In addition to guaranteeing complete coverage of the parameter ranges described in Sec.~\ref{sec:requirements for parameter coverage}, we define a single distribution that spans the whole space rather than several \textit{ad hoc} subpopulations, as was done in O3 \cite{GWTC-3-injections, GWTC-3-cumulative-injections}.
The distribution should have broad tails (to support analyses of possible outliers) and be relatively simple so that it is easy to reproduce.
This includes choosing a distribution that can be factored into separate one-dimensional distributions (e.g., $p(m_1, m_2, \vec{s}_1, \vec{s}_2, z) = p(m_1) p(m_2|m_1) p(\vec{s}_1) p(\vec{s}_2) p(z)$) so there are no complicated correlations between parameters in the injected distribution.
The only exception to this is $p(m_2|m_1, \Lambda_\mathrm{inj})$, which includes the convention $m_1 \geq m_2$ and therefore has an upper limit that depends on $m_1$.

Additionally, search teams requested that injections be divided into separate sets of similar detector-frame chirp mass (the best-measured parameter for most CBCs) and that injections be spaced at least 24 sec apart.
Sec.~\ref{sec:injection generation workflow} describes how this was done by reordering independently and identically distributed (i.i.d.) injections drawn from a single distribution to construct a ``checkerboard'' pattern in mass among four subsets.

We also included a few quantitative sanity checks.
Specifically, we estimated the $N_\mathrm{eff}$ that would be obtained when reweighting the injections to follow several small, targeted distributions with the goal of guaranteeing that we would have enough injections to resolve features of interest.
This included possible peaks in the source-frame primary mass distribution near 10, 25, and $35\,M_\odot$.
We also tested distributions that favor smalls spin magnitudes and (mostly) aligned spin orientations.
As a final sanity check, we computed $N_\mathrm{eff}$ throughout the prior range for the \textsc{PowerLaw+Peak} model used in O3~\cite{GWTC-3-RnP}.

Although we did not place any strict requirements on the exact $N_\mathrm{eff}$ for any of these targeted populations, we did use them as basic sanity checks to guide the iterative design process described in Sec.~\ref{sec:iterative design}.


\subsection{Iterative Design}
\label{sec:iterative design}

Starting from an initial proposal, which was described by
\begin{itemize}
    \item a broken powerlaw in $m_1^{(\mathrm{src})}$ that roughly follows the key features observed in O3 and a long tail to high masses with
        \begin{itemize}
            \item a high rate below $\sim 3\, M_\odot$,
            \item a steep drop between $\sim 3$-$10\, M_\odot$,
            \item a flat(-ish) distribution between 10 and $50\, M_\odot$, and
            \item a steepening near $\sim 50\,M_\odot$;
        \end{itemize}
    \item a broad powerlaw in asymmetric mass ratio $q = m_2/m_1$;
    \item broad support for spin magnitudes and tilts;
    \item a local merger rate that increases with redshift; 
    \item isotropic distributions over inclination, right ascension, declination, and other orientation parameters,
\end{itemize}
we iteratively evaluated its performance and make changes by
\begin{itemize}
    \item generating a semianalytic sensitivity estimate (set of detected injections) with realistic selection criteria~\cite{Essick:2023toz},
    \item computing $N_\mathrm{eff}$ for our targeted set of subpopulations, and
    \item checking to make sure that $N_{\mathrm{eff}}$ is large enough compared to the expected number of detections from that subpopulation.
\end{itemize}
The proposed distribution was then updated in order to address any shortcomings.

Finally, we additionally checked that the ratio of the number of detections for binary neutron star (BNS: $m_2^{(\mathrm{src})} \leq m_1^{(\mathrm{src})} \leq 3\,M_\odot$), neutron star-black hole (NSBH: $m_2^{(\mathrm{src})} \leq 3\,M_\odot \leq m_1^{(\mathrm{src})}$), and binary black hole (BBH: $3\,M_\odot \leq m_2^{(\mathrm{src})} \leq m_1^{(\mathrm{src})} \leq 100\,M_\odot$) systems approximately matches what was observed during previous runs \cite{GWTC-3}: \result{$O(2\%)$ BNS, $O(2\%)$ NSBH, and $O(96\%)$ BBH}.
The distribution described in Sec.~\ref{sec:distribution} yields approximately \result{1\% BNS, 1\% NSBH, and 93\% BBH with the remaining 5\% corresponding to IMBH ($m_1^{(\mathrm{src})} > 100\,M_\odot$)}.


\subsection{Injected Distribution}
\label{sec:distribution}

\begin{figure*}
    \begin{minipage}{1.0\columnwidth}
        \includegraphics[width=1.0\columnwidth]{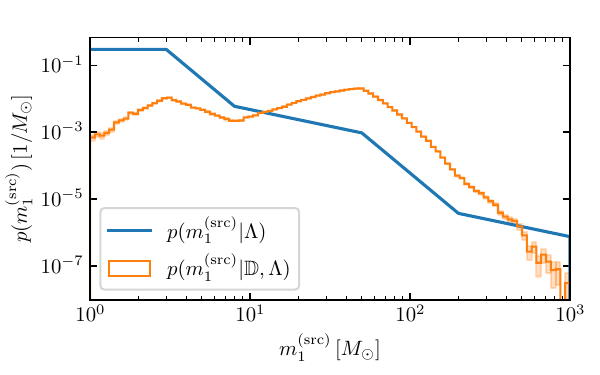}
        \includegraphics[width=1.0\columnwidth]{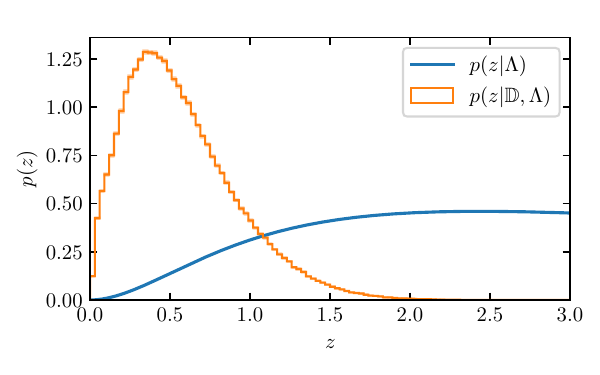}
    \end{minipage}
    \begin{minipage}{1.0\columnwidth}
        \includegraphics[width=1.0\columnwidth]{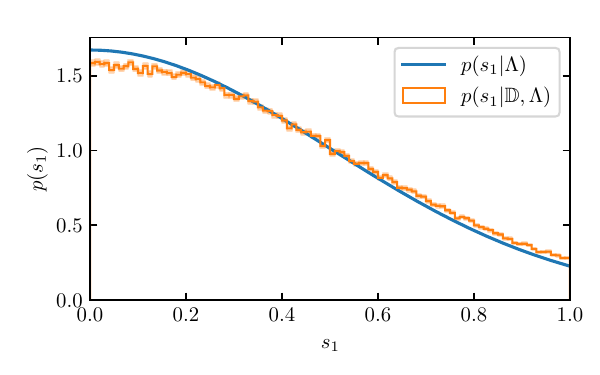}
        \includegraphics[width=1.0\columnwidth]{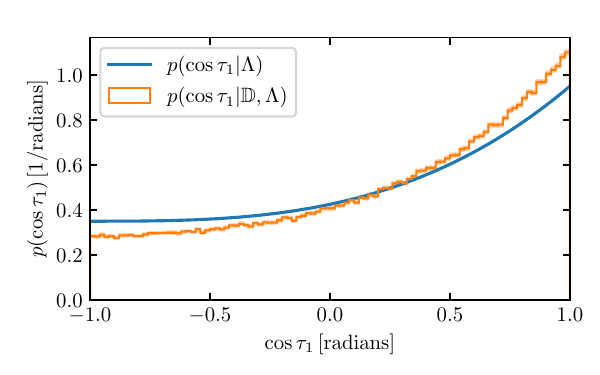}
    \end{minipage}
    \caption{
        Injected (\emph{blue}) and recovered (\emph{orange}) distributions for the O4a injection campaign for the (\emph{top left}) source-frame primary mass, (\emph{bottom left}) redshift, and primary spin (\emph{top right}) magnitude and (\emph{bottom right}) tilt or polar angle.
        An event is detected if at least one search reported \result{FAR $< 1$/year}.
        Shaded bands for $p(\cdot|\mathbb{D},\Lambda)$ represents 1-$\sigma$ uncertainty from the finite number of samples.
    }
    \label{fig:distribution}
\end{figure*}

\begin{figure*}
    \includegraphics[width=1.0\textwidth]{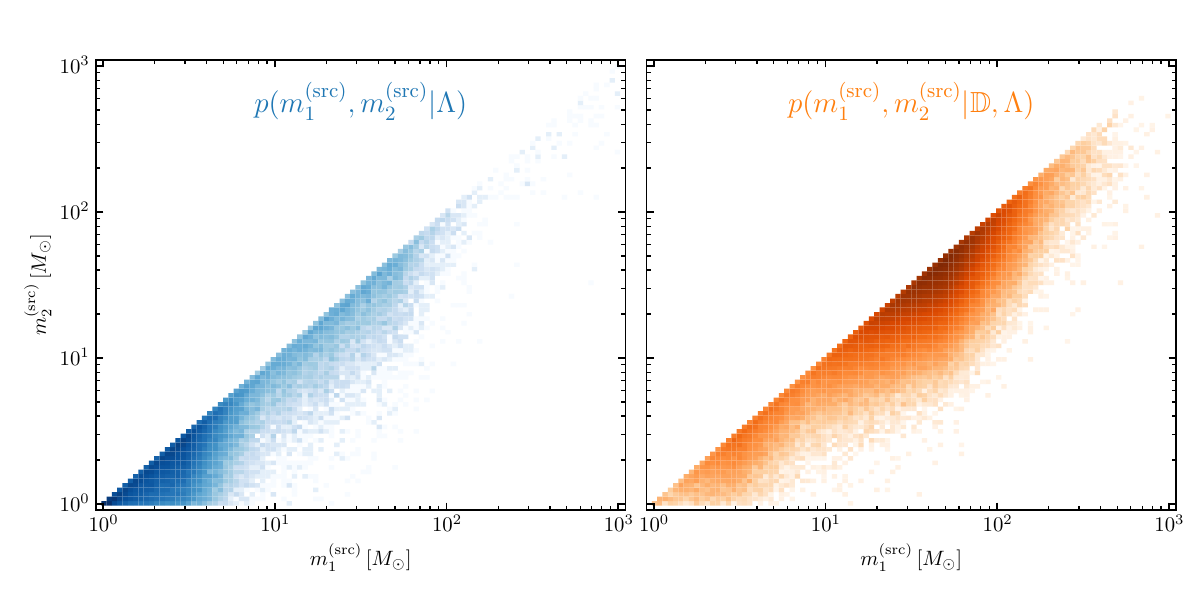}
    \caption{
        (\emph{left}) Injected and (\emph{right}) recovered joint distributions over source-frame component masses.
        An event is detected if at least one search reported \result{FAR $< 1$/year}.
        Shading corresponds to logarithmic scaling for the probability density.
    }
    \label{fig:joint mass distribution}
\end{figure*}

\begin{figure*}
    \includegraphics[width=1.0\textwidth]{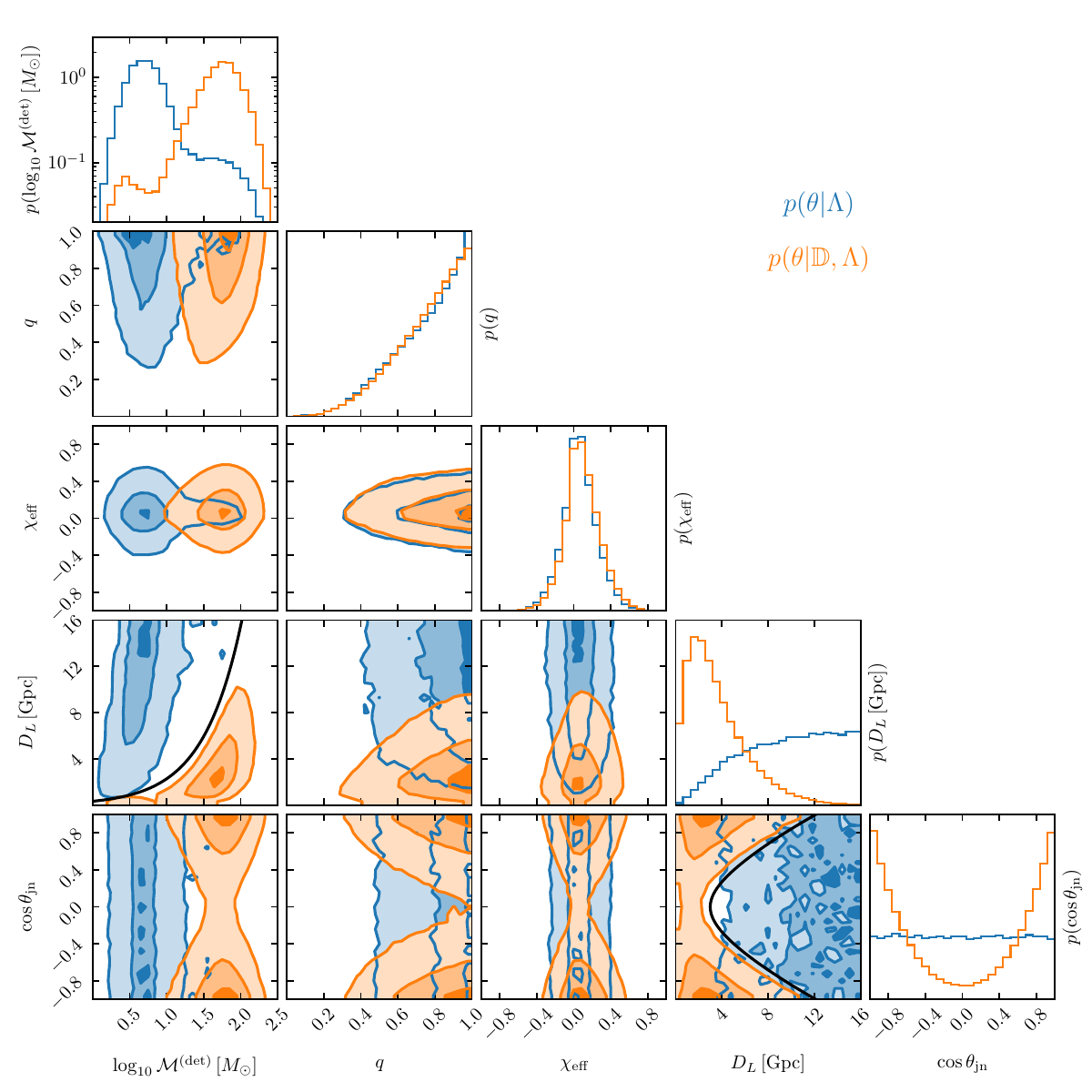}
    \caption{
        Joint and marginal distributions for the (\emph{blue}) injected and (\emph{orange}) recovered distributions over (\emph{left to right and top to bottom}) detector-frame chirp mass, asymmetric mass ratio, effective inspiral spin, luminosity distance, and inclination.
        Events are detected if at least one search reported \result{FAR $< 1$/year}.
        We additionally annotate some joint distributions with (\emph{black lines}) exclusion regions based on expected scaling of the signal amplitude: $D_L \propto (\mathcal{M}^{(\mathrm{det})})^{5/6}$ and $D_L \sim (\cos^2\theta_{jn} + ((1+\cos^2\theta_{jn})/2)^2)^{1/2}$.
    }
    \label{fig:corner}
\end{figure*}

\begin{figure}
    \includegraphics[width=1.0\columnwidth]{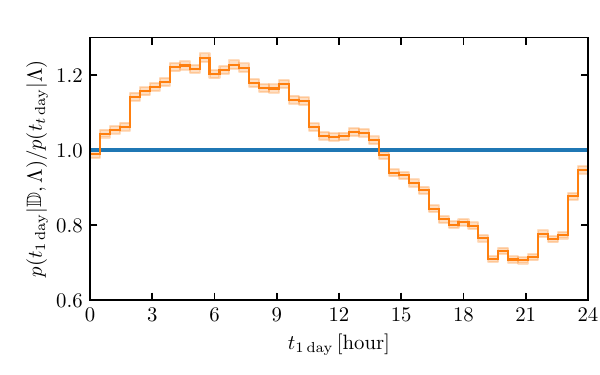}
    \includegraphics[width=1.0\columnwidth]{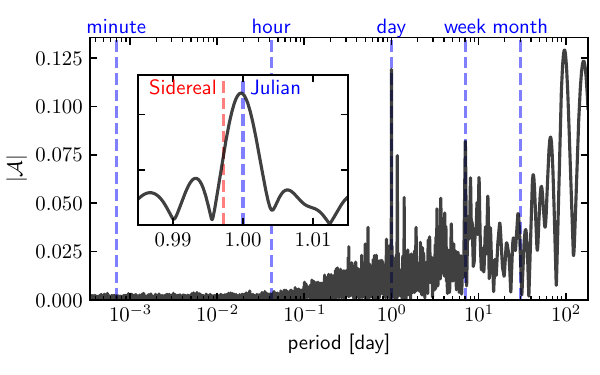}
    \caption{
        (\emph{top}) A folded histogram of the time-of-day (GPS time mod 86400 sec) for (\emph{blue}) injected and (\emph{orange}) recovered events, showing a clear diurnal cycle of comparable magnitude to what was observed in O1~\cite{Chen:2016luc}.
        Midnight UTC roughly corresponds to zero in these units, so the peak near \result{$\sim 6$ hours} corresponds to the middle of the night in North America.
        Events are detected if at least one search reported \result{FAR $< 1$/year}.
        Shaded band represents 1-$\sigma$ uncertainty from the finite number of events.
        (\emph{bottom}) A Fourier transform of $p(t|\mathbb{D},\Lambda_\mathrm{inj})$ (Eq.~\ref{eq:fancy fourier}) showing strong periodic signals at one day (diurnal cycle) and one week (regular Tuesday maintenance).
        The inset provides a closer view of the feature at one day.
    }
    \label{fig:time}
\end{figure}

\begin{figure}
    \includegraphics[width=1.0\columnwidth]{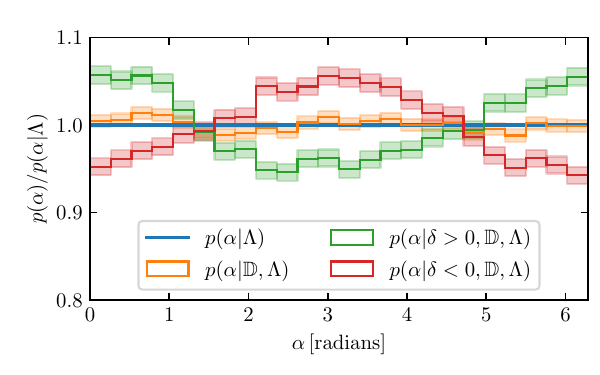}
    \includegraphics[width=1.0\columnwidth]{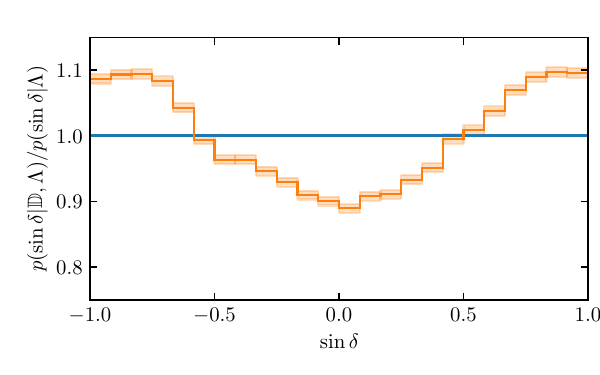}
    \caption{
        Injected (\emph{blue}) and recovered (\emph{orange}) distributions for the (\emph{top}) right ascension ($\alpha$) and (\emph{bottom}) declination ($\delta$) of sources.
        Events are detected if at least one search reported \result{FAR $< 1$/year}.
        We additionally show distributions of $\alpha$ from the northern ($\delta > 0$) and southern ($\delta < 0$) hemispheres separately to highlight the selection introduced by diurnal cycles within the detector network (Fig.~\ref{fig:time}).
        Shaded bands represent 1-$\sigma$ uncertainty from the finite number of samples.
    }
    \label{fig:sky loc}
\end{figure}

We now describe the injection distribution used throughout O4 and compare the injected and recovered distributions in Figs.~\ref{fig:distribution}--\ref{fig:sky loc}.
While we provide the details of the injected distribution used, we also record the draw probabilities $p(\theta_e|\Lambda_\mathrm{inj})$ alongside the parameters of individual injections ($\theta_e$) directly within the public data products~\cite{GWTC-4-cumulative-injections, GWTC-4-injections}.
As such, users do not need to recompute these themselves.

We define a joint distribution $p(\theta|\Lambda_\mathrm{inj})$ as the product of the following.
The source-frame masses are distributed according to a five-piece broken power-law for the primary mass between 1--$10^3\,M_\odot$ and a simple power law for the secondary mass (conditioned on the primary mass) from $1\,M_\odot$ to $m_1^{(\mathrm{src})}$:
\begin{widetext}
\result{
\begin{equation}
    p(m_1^{(\mathrm{src})}|\Lambda_\mathrm{inj}) \propto
\left\{ \begin{array}{ll}
    0.0                                                                                                              & m_1^{(\mathrm{src})} < 1 \\
    1.0                                                                                                              & 1 \leq m_1^{(\mathrm{src})} < 3 \\
    \left(m_1^{(\mathrm{src})}/3\right)^{-4}                                                                         & 3 \leq m_1^{(\mathrm{src})} < 8 \\    
    \left(8/3\right)^{-4} \left(m_1^{(\mathrm{src})}/8\right)^{-1}                                                   & 7 \leq m_1^{(\mathrm{src})} < 50 \\
    \left(8/3\right)^{-4} \left(50/8\right)^{-1} \left(m_1^{(\mathrm{src})}/50\right)^{-4}                           & 50 \leq m_1^{(\mathrm{src})} < 200 \\
    \left(8/3\right)^{-4} \left(50/8\right)^{-1} \left(200/50\right)^{-4} \left(m_1^{(\mathrm{src})}/200\right)^{-1} & 200 \leq m_1^{(\mathrm{src})} < 1000 \\
    0.0                                                                                                              & 1000 < m_1^{(\mathrm{src})} \\
\end{array} \right.
\end{equation}
}
\end{widetext}
and
\result{
\begin{equation}
    p(m_2^{(\mathrm{src})}|m_1^{(\mathrm{src})}, \Lambda_\mathrm{inj}) = \frac{2 m_2^{(\mathrm{src})}}{(m_1^{(\mathrm{src})})^2 - 1} \Theta(1 \leq m_2^{(\mathrm{src})} \leq m_1^{(\mathrm{src})})
\end{equation}
}

Component spins are i.i.d. with a truncated Gaussian for the spin magnitude ($s$) centered at zero with a standard deviation of 0.5
\result{
\begin{equation}
    p(s|\Lambda_\mathrm{inj}) \propto e^{-2s^2} \ , 
\end{equation}
}
a mixture of aligned ($\beta$-distributed) and isotropic spin tilts ($\tau$)
\result{
\begin{align}
    p(\cos\tau|\Lambda_\mathrm{inj})
        & = 0.3 \left(\frac{(1+\cos\tau)^3}{4}\right) + 0.7\left(\frac{1}{2}\right) \ ,
\end{align}
}
and uniformly distributed azimuthal angles: \result{$\phi \in [0, 2\pi)$}.
Spins are defined at a reference frequency of \result{16 Hz}.

\result{The redshift is distribution grows with redshift faster than a uniform distribution in co-moving volume ($V_c$) and source frame time}
\begin{equation}
    p(z|\Lambda_\mathrm{inj}) \propto \frac{dV_c}{dz} \Theta(0 \leq z < 3)
\end{equation}
or, equivalently, a local merger rate that grows with redshift
\begin{equation}
    \frac{dN}{dV_c dt_\mathrm{src}} \propto (1+z)
\end{equation}
assuming a flat $\Lambda$CDM cosmology with \result{$H_0 = 67.90\ \mathrm{km/s/Mpc}$, $\Omega_m = 0.3065$, and $\Omega_\Lambda = 1 - \Omega_m$} \cite{Planck:2015fie}.
We additionally assume an isotropic distribution of inclinations (\result{$\cos \theta_\mathrm{jn} \in [-1, +1]$}), uniform distributions over polarization angle (\result{$\psi \in [0, \pi)$}) and the phase at coalescence (\result{$\phi_c \in [0, 2\pi)$}), and isotropic distributions over the right ascension ($\alpha$) and declination ($\delta$): \result{$\alpha \in [0, 2\pi)$} and \result{$\sin\delta \in [-1, +1]$}.
As mentioned in Sec.~\ref{sec:requirements for parameter coverage}, we assume all systems are \result{quasi-circular (zero eccentricity)} and \result{do not include tidal effects}.
We additionally distribute signals with a spacing of six seconds with an additional random offset uniformly distributed within \result{$[-0.5, +0.5]\,\mathrm{sec}$}.

Figs.~\ref{fig:distribution} and~\ref{fig:joint mass distribution} show the non-trivial injected distributions over the astrophysically relevant parameters, such as the source-frame masses, redshift, and spins as well as the distribution recovered by conditioning on at least one search reporting \result{FAR $< 1$/year}.
Complementarily, Fig.~\ref{fig:corner} again shows the injected and recovered distributions but in coordinates that are more relevant for the CBC waveform, such as the detector-frame chirp mass
\begin{equation}
    \mathcal{M}^{(\mathrm{det})} = (1+z) \frac{\left(m_1^{(\mathrm{src})} m_2^{(\mathrm{src})}\right)^{3/5}}{\left(m_1^{(\mathrm{src})} + m_2^{(\mathrm{src})}\right)^{1/5}} \ ,
\end{equation}
the effective inspiral spin
\begin{equation}
    \chi_\mathrm{eff} = \frac{m_1^{(\mathrm{src})} s_1 \cos \tau_1 + m_2^{(\mathrm{src})} s_2 \cos\tau_2}{m_1^{(\mathrm{src})} + m_2^{(\mathrm{src})}} \ ,
\end{equation}
the luminosity distance ($D_L$), and the orbital inclination ($\theta_\mathrm{jn}$) between the orbital angular momentum and the line-of-sight to the source (again, defined at the reference frequency of \result{16 Hz}).
We can clearly see correlations induced by the selection between $D_L$ and $\mathcal{M}^{(\mathrm{det})}$ as well as $D_L$ and $\cos\theta_\mathrm{jn}$ driven by the combinations of these parameters that affect the GW amplitude at leading order.
Specifically, we expect the frequency domain signal amplitude in the $+$ and $\times$ polarizations to scale as
\begin{equation}
    \begin{bmatrix} h_+ \\ h_\times \end{bmatrix} \propto \frac{(\mathcal{M}^{(\mathrm{det})})^{5/6}}{D_L} \begin{bmatrix} (1+\cos^2\theta_\mathrm{jn})/2 \\ \cos\theta_\mathrm{jn} \end{bmatrix}
\end{equation}
We see only a weak selection on $\chi_\mathrm{eff}$, slightly favoring larger values because of the orbital hang-up effect~\cite{Ng:2018neg}, and essentially no selection on the asymmetric mass ratio $q = m_2 / m_1$.

The selections on mass, distance, inclination, and effective spin are all relatively well-known and come from the physics of CBC systems themselves.
However, there are other terrestrial effects evident in the detected distribution.
Specifically, Fig.~\ref{fig:time} shows a folded histogram for the coalescence time, i.e., the coalescence time mod one day (\externalresult{86400 sec}).
There is a clear diurnal sinusoidal pattern within the recovered distribution $p(t|\mathbb{D},\Lambda)$ with an amplitude of $\gtrsim 20\%$ relative to the uniform injected distribution.
This is comparable in magnitude to the diurnal cycle observed during O1~\cite{Chen:2016luc}.
Although not fully explained, this is thought to be due to anthropogenic noise (i.e., human activity) at the sites.
Fig.~\ref{fig:time} also shows the amplitude of Fourier coefficients as a function of period ($p$) estimated via
\begin{align}
    \mathcal{A}
        & = \int dt\, p(t|\mathbb{D},\Lambda_\mathrm{inj}) e^{-2\pi i t/p} \nonumber \\
        & \approx \frac{1}{N_\mathrm{fnd}} \sum\limits_e^{N_\mathrm{fnd}} e^{-2\pi i t_e/p} \label{eq:fancy fourier}
\end{align}
There are strong periodic components at \result{one day (diurnal cycles)} and \result{one week (regular maintenance on Tuesdays)}.
The peak at one day (\externalresult{86400 sec}) is narrow enough to clearly distinguish it from the sidereal day (\externalresult{86164.0905 sec}).
There are additional unexplained periodic components at \result{$\sim1.16$} and \result{$\sim1.40$} days as well as broadband structure at very long periods (\result{$\gtrsim$ month}).
The latter are likely associated with detector commissioning and/or secular sensitivity changes throughout the run.

Finally, Fig.~\ref{fig:sky loc} shows the induced selection over declination ($\delta$) relative to the injected distribution, caused by the detectors' locations at mid-latitudes in the northern hemisphere and the shape of the antenna patterns, along with the selection over right ascension ($\alpha$) relative to the injected distribution.
The non-trivial selection over $\alpha$ is caused by the nontrivial selection over the events' time at the same frequency as the Earth's rotation, and it is reversed in the northern and southern hemispheres.
This is expected from the quadrupolar nature of the GW detectors' antenna patterns and the fact that the network sensitivity is dominated by the two (nearly aligned) LIGO interferometers.
As discussed in~\citet{Chen:2016luc}, these types of effects can affect the feasibility of prompt multi-messenger follow-up of GW events.
They are also important when constraining the (an)isotropy of the merger rate density~\cite{Essick:2022slj, Stiskalek:2020wbj, Payne:2020pmc}.

\subsubsection{Weaknesses in the Injected Distribution}
\label{sec:weaknesses}

In general, the current injection set satisfies the design requirements reasonably well.
However, it is known that this injection distribution favors certain types of binaries.
This was an intentional design choice, as the injected distribution should closely follow the astrophysical distribution to maximize the statistical precision available through importance sampling~\cite{Essick:2022ojx}.

More generally, we note that the injected distribution has
\begin{itemize}
    \item few effective samples for component mass distributions with extremely large primary and small secondary masses (i.e., with very asymmetric mass ratios), although there is good coverage overall in the asymmetric mass ratio for $q > 0.1$ (Fig.~\ref{fig:corner}),
    \item few samples near $\chi_\mathrm{eff} \sim \pm 1$ (most of the support is for $|\chi_\mathrm{eff}| \lesssim 0.5$),
    \item broad support near very tightly (anti)aligned spin configurations ($|\cos\tau| \sim 1$) and small spin magnitudes ($s \sim 0$); narrow population models may suffer from significant Monte Carlo uncertainty.
\end{itemize}


\section{Injection Generation Workflow}
\label{sec:injection generation workflow}

Having established the injection distribution for O4, we now briefly document the workflow used to generate production tabular and timeseries data.
In order to make injections available promptly, separate injection sets were created at the end of each month.
The workflow was divided into several steps: PSD estimation, drawing injections, and generating timeseries data for searches.
Searches then process the data and report detection statistics and signal reconstructions.
Once all search data was collected, we created a mixture model~\cite{Essick:2021} spanning all of O4a from the separate month-by-month injection sets.


\subsection{Estimating PSDs}
\label{sec:estimating psds}

PSDs were estimated using Welch's algorithm \cite{Welch:1967oth} every 60 seconds (15 segments, each 4 seconds long with zero overlap, using a Tukey window with $\alpha=0.5$).
Science time was defined separately in each IFO by the flags \texttt{H1:DMT-ANALYSIS\_READY}, \texttt{L1:DMT-ANALYSIS\_READY}, and \texttt{V1:ITF\_SCIENCE} for LHO, LLO, and Virgo, respectively.
The GPS start time of each segment was an integer multiple of 60 seconds, and we only estimated PSDs if the entire 60 sec window was contained within the science segments for that IFO.

Given an ensemble of PSDs within a month, we then took the 10\% quantile in each frequency bin for each IFO separately as the reference value used to generate injections for that month.
We chose the 10\% quantile to be conservative (slightly overestimate the actual range).
However, this does not affect any of the real search results as the reference PSD is only used to estimate $\rho_\mathrm{opt,net}$ when applying the hopeless cut~\cite{Essick:2023toz}.

Fig.~\ref{fig:psd} shows the nine reference PSDs estimated for each of LHO and LLO during O4a (Virgo did not participate in O4a) along with the expected PSD predicted before the beginning of O4a.
There is some noticeable variability from month to month, and the actual O4a sensitivities did not quite reach the projected sensitivity, particularly between 20--50 Hz.


\subsection{Drawing Injections}
\label{sec:drawing injections}

\begin{figure*}
    \includegraphics[width=1.0\textwidth]{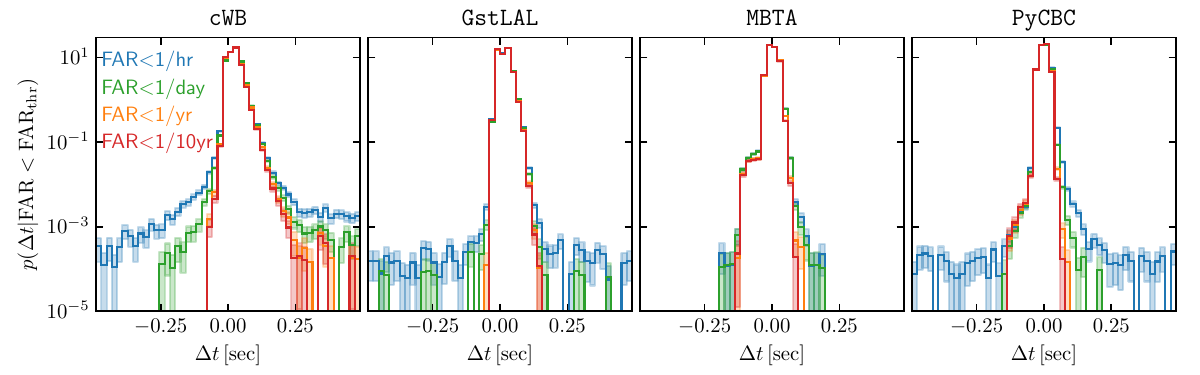}
    \caption{
        Distributions of the difference between times reported by searches and the reference injection time at geocenter ($\Delta t = t_\mathrm{rec} - t_\mathrm{inj}$) for (\emph{left to right}) \texttt{cWB}, \texttt{GstLAL}, \texttt{MBTA}, and \texttt{PyCBC} for event selected with different FAR thresholds: (\emph{blue}) $<$ 1/hour, (\emph{green}) $<$ 1/day, (\emph{orange}) $<$ 1/year, and (\emph{red}) $<$ 1/10 years.
        Shaded bands represent 1-$\sigma$ uncertainty from the finite number of samples.
        Although not readily apparent here, systematic offsets in recovered times may be associated with differences in conventions used in different waveform approximants.
        However, the most physically relevant effect is the time delay between IFOs, which informs source localization and is not affected by such systematic time offsets.
    }
    \label{fig:search times}
\end{figure*}

We then drew a large set of injections, enough to place one every six seconds in detector-frame wall-time.
This was accomplished by drawing samples from the injected distribution $p(\theta|\Lambda_\mathrm{inj})$, computing the (expected) $\rho_\mathrm{opt,net}$ using reference PSDs, and retaining only those events that passed the hopeless cut: \result{$\rho_\mathrm{opt,net} > \rho_\mathrm{cut} = 5$}.
Only the events that pass the hopeless cut were actually passed to searches.
During O4a, we drew \result{433,157,717} events from $p(\theta|\Lambda_\mathrm{inj})$ but only passed \result{3,974,400} events to searches.

Additionally, the entire injection set for searches was first drawn separately for each month.
Within each month, these injections were drawn at a fixed GPS time (equivalent to drawing them in a frame fixed to the Earth).
After the i.i.d. events were obtained, we then divided them by $\mathcal{M}^{(\mathrm{det})}$ into four subsets of equal number.
Events were then randomly selected from each subset (without replacement) and ordered in time with a regular spacing of \result{6 sec} with up to $\pm 0.5$ sec of random jitter (drawn from a uniform distribution around the regular spacing), taking care to update the right ascension of sources with the GPS time (at geocenter) to maintain the same relative orientation between the source and the detectors.
We placed events sequentially from most massive (shortest waveform) to least massive (longest waveforms). 
That is, for subsets A, B, C, and D, we place events in the following order: A B C D A B C D A ... so that there were \result{6 sec} between nearest neighbors (A to B and B to C, etc) and \result{24 sec} between events drawn from the same subset (A to A and B to B, etc).
This pattern was repeated through the entire detector-frame wall-time for each month.\footnote{The duration of each monthly injection set was chosen to be an \result{integer multiple of 24 sec}. Consequently, the injection sequences were seamless between months.}
IFO duty cycles are therefore automatically encoded within the times of the events that are detected by searches.

For each subset separately, we then supplemented the tabular data ($\theta_e$ and $p(\theta_e|\Lambda_\mathrm{inj})$) with injection-only timeseries of projected strains in each IFO.
While there could be some overlap in time between neighboring signals, particularly at low masses, this did not affect searches' ability to distinguish between them because their time-frequency tracks overlap only minimally.
Separate timeseries frames for each subset of injections were then passed to searches for further processing.


\subsection{Searches}
\label{sec:searches}

\begin{figure*}
    \includegraphics[width=1.0\textwidth]{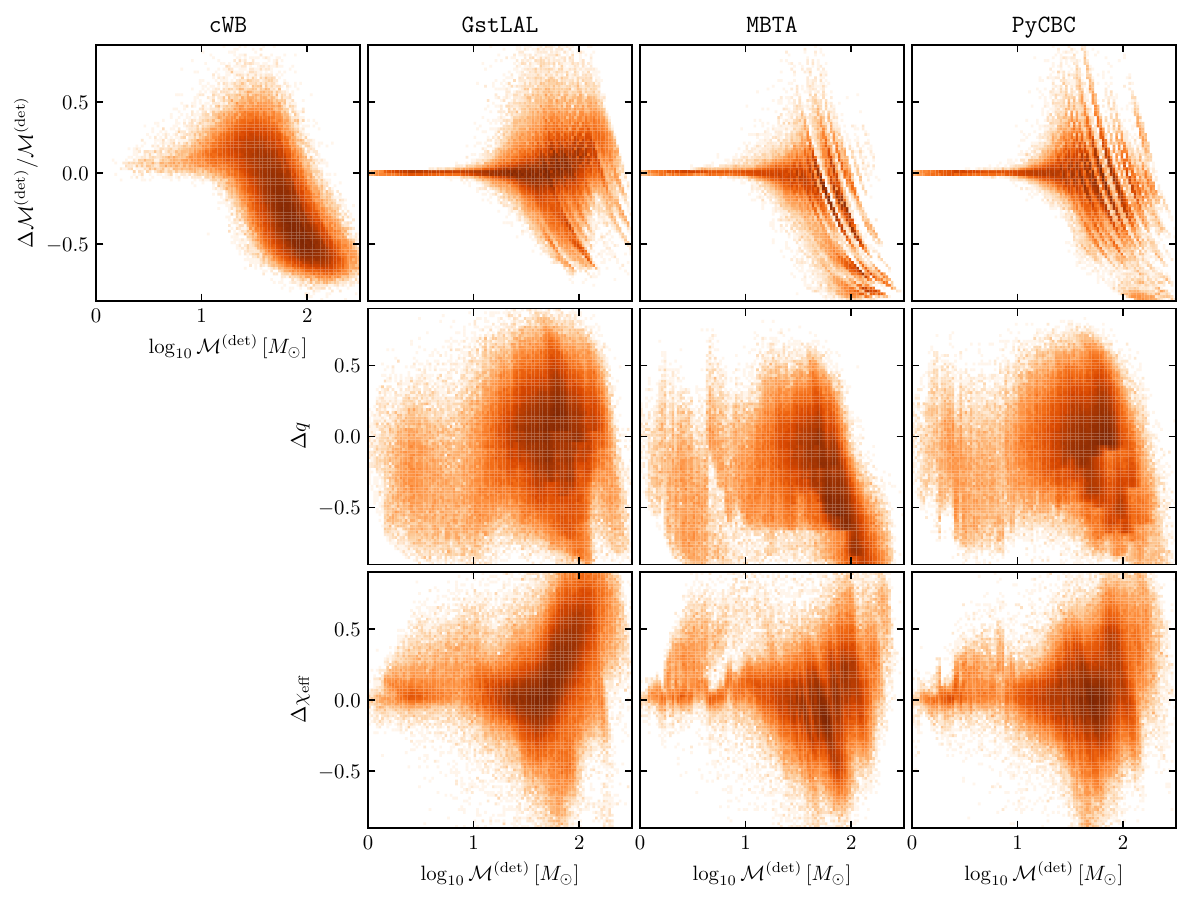}
    \caption{
        Distributions of differences between reconstructed parameters and injected parameters for (\emph{left to right}) \texttt{cWB}, \texttt{GstLAL}, \texttt{MBTA}, and \texttt{PyCBC} for events detected by each search separately with \result{FAR $<$ 1/year}.
        We show the (\emph{top row}) fractional error in the detector-frame chirp mass, (\emph{middle row}) the error in the asymmetric mass ratio, and (\emph{bottom row}) the error in the effective inspiral spin.
        Differences are between the recovered and injected parameters (i.e., $\Delta X = X_\mathrm{rec} - X_\mathrm{inj}$).
        Shading corresponds to a logarithmic scale for probability densities.
    }
    \label{fig:search masses}
\end{figure*}

\begin{figure*}
    \includegraphics[width=1.0\textwidth]{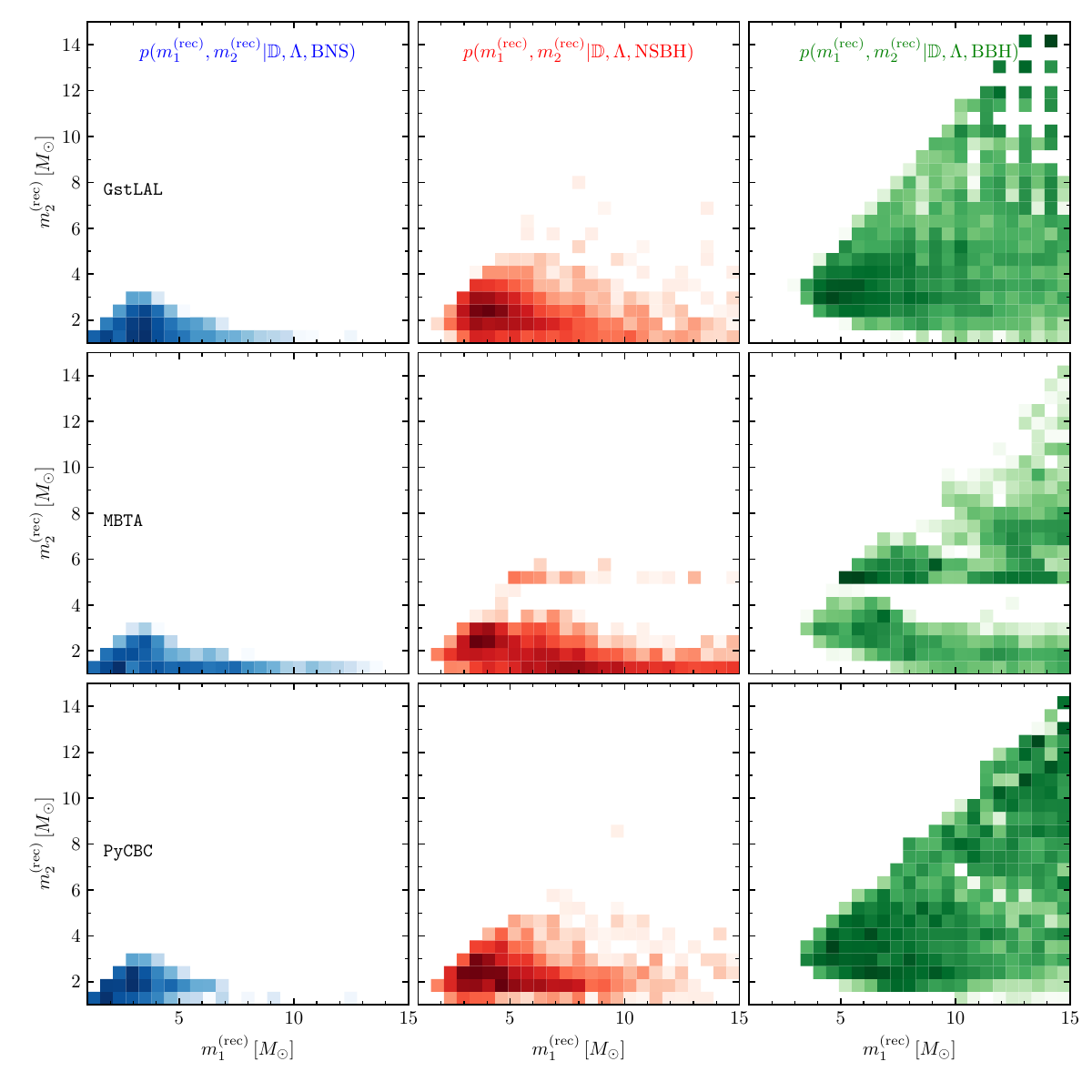}
    \caption{
        Distributions of recovered detector-frame component masses for (\emph{top}) \texttt{GstLAL}, (\emph{middle}) \texttt{MBTA}, and (\emph{bottom}) \texttt{PyCBC} for injections classified at (\emph{left, blue}) \result{BNS ($m_2^{(\mathrm{src})} \leq m_1^{(\mathrm{src})} \leq 3\,M_\odot$)}, (\emph{middle, red}) \result{NSBH ($m_2^{(\mathrm{src})} \leq 3\,M_\odot \leq m_1^{(\mathrm{src})}$)}, and (\emph{right, green}) \result{BBH ($3\,M_\odot \leq m_2^{(\mathrm{src})} \leq m_1^{(\mathrm{src})}$)}.
        Shading corresponds to a logarithmic scale for probability density.
        As in Fig.~\ref{fig:search masses}, we see a tendency for searches to recover template masses that are significantly more asymmetric than the true masses.
    }
    \label{fig:search categories}
\end{figure*}

During O4, four searches processed offline injections: coherent waveburst (\texttt{cWB}; \cite{Klimenko:2011, Klimenko:2004, Klimenko:2015, Mishra:2025}), \texttt{GstLAL} (\cite{Tsukada:2023edh, Messick:2017, Sachdev:2019, Hanna:2019, Cannon:2021}), Multi-Band Template Analysis (\texttt{MBTA}; \cite{Allene:2025saz, Adams:2015, Aubin:2020}), and \texttt{PyCBC} (\cite{Allen:2004, DalCanton:2014, Usman:2015, Nitz:2017, Davies:2020, Allen:2005}).
\texttt{GstLAL}, \texttt{MBTA}, and \texttt{PyCBC} are all template-based searches that run data through matched-filters spanning large template banks of theoretical CBC waveforms.
\texttt{cWB}, on the other hand, does not assume a template but instead coherently reconstructs signals in the wavelet domain through a constrained/regularized maximum-likelihood approach.
Different versions of \texttt{cWB} impose different regularization schemes to target different types of signals.

Each of the four searches returned statistics for all the events they found within O4.
Specifically, they all provided FARs, ranking statistics, and recovered signal-to-noise ratios in each IFO and over the entire network.
\texttt{cWB}, \texttt{GstLAL}, and \texttt{MBTA} additionally provided probabilities that the sources were astrophysical in origin ($p_\mathrm{astro}$~\cite{GWTC-3}).
Template-based searches provided the (detector-frame) mass and spin parameters of the best-match template for each signal.
\texttt{cWB} instead provided moments of the time and frequency of the signal (with respect to reconstructed signal power) as well as an estimate of $\mathcal{M}^{(\mathrm{det})}$.

Events reported by searches were matched with injections as follows.
First, any event reported by any search that is within \result{$\pm 0.25$ sec} of any real candidate found in the GWTC-4 catalog (again, by any search) was removed.
This is necessary because we inject into real detector data, which also contains real GWs.
The number of events removed is negligibly small compared the total size of the injection set.
After that, each remaining search event is matched with the closest injection in time.
In the case of multiple matches from the same search to the same injection, we retain the search event with the smallest FAR.
Fig.~\ref{fig:search times} shows the distribution of the differences between the times reported by searches and the injection time (at geocenter) for several FAR thresholds.
Fig.~\ref{fig:search masses} shows other reconstructed vs. injected parameters for events detected with \result{FAR $\leq 1/\mathrm{year}$}.
Fig.~\ref{fig:search categories} shows distributions of best-match template masses conditioned on the injected source-frame masses, again for \result{FAR $\leq 1/\mathrm{year}$}, from which low-latency source-classification can be derived \cite{Chatterjee:2019avs, Chaudhary:2023vec}.

Figs.~\ref{fig:search times},~\ref{fig:search masses}, and~\ref{fig:search categories} make it clear that the behavior of search pipelines can be very complicated, but this behavior can nevertheless be modeled using the public data products~\cite{GWTC-4-cumulative-injections, GWTC-4-injections}.


\subsection{Constructing Mixture Models}
\label{sec:mixture model construction}

Once we obtained all search results, we combined the monthly injection sets into a single combined mixture model spanning all of O4a following~\citet{Essick:2021}.
This procedure \textit{de facto} reweights each injection to make it appear as if they were injected at the same rate.
Specifically, because the reference PSDs change from month to month, the acceptance fraction (ratio of trials that pass the hopeless cut to the total number of trials) also changes, and the total number of trials follows an negative-binomial distribution.
Because injections in different months are therefore not i.i.d., we must take care when combining them.
The resulting data product~\cite{GWTC-4-cumulative-injections, GWTC-4-injections} contains an additional field (\texttt{weights}) that captures this reweighting process.
Again, see~\citet{Essick:2021} for more details.

We also combined the O4a mixture model with other sensitivity estimates for O1, O2, and O3 to form a cumulative mixture model spanning O1 -- O4a.
We generated new semianalytic injection sets for O1 and O2 separately while we used the previously released real injections from O3~\cite{GWTC-3-injections}.

The new data products are therefore released as ``O4a-only''~\cite{GWTC-4-injections} and as cumulative ``O1+O2+O3+O4a'' mixtures~\cite{GWTC-4-cumulative-injections}.


\section{Discussion}
\label{sec:discussion}

We have documented the design process, injected distribution, and basic results for CBC sensitivity estimates using real LVK searches and GW data during O4.
In addition to the expected correlations induced by selection effects, we clearly identified subtler effects in the distribution of detector-frame time and source location on the celestial sphere.
We additionally showed distributions of signal reconstructions returned by searches, which can be used to inform low-latency classification and follow-up.

The campaign for O4a (\result{May 2023 -- January 2024}) produced \result{476,415} detected injections (at least one search reported \result{FAR$<1$/year}).
Results are publicly available within Zenodo~\cite{GWTC-4-cumulative-injections, GWTC-4-injections}.
Campaigns of similar scope are underway for O4b (\result{March 2024 -- January 2025}) and O4c (\result{January 2025 -- November 2025}).
All told, we expect to produce \result{$> 10^6$} detected injections for all of O4, which should provide sufficiently precise estimates of $P(\mathbb{D}|\Lambda)$, for reasonable populations, to accommodate catalogs of \result{$O(500)$} events or more.
Additional public data products will be released alongside future catalogs.

Although the new data products~\cite{GWTC-4-cumulative-injections, GWTC-4-injections} represent a significant improvement on previous efforts to characterize sensitivity estimates by introducing a larger sample size and more carefully chosen injected distribution (see additional discussion in Sec.~\ref{sec:introduction}), we remind the reader that the public data products are not completely foolproof.

In particular, users still need to explicitly check that Monte Carlo approximations to high-dimensional integrals are still reliable.
This is often implemented through criteria for the ``effective number of samples'' ($N_\mathrm{eff}$~\cite{Farr:2019rap}).
See, e.g.,~\citet{Essick:2022ojx} for a broader discussion and suggestions.

Additionally, cumulative data products like Ref.~\cite{GWTC-4-cumulative-injections} inherit choices made for previous runs.
Even though the injected distribution for O4 covers a wide range of parameter space and the updated semianalytic samples for O1 and O2 follow the same distribution (except for the maximum redshift), the cumulative estimates still contain ``holes'' in the parameter space from the injected distributions used during O3~\cite{GWTC-3-injections, GWTC-3-cumulative-injections}.
In particular, there are no low-mass ($m^{(\mathrm{src})} < 2\,M_\odot$) and high-spin ($|s| > 0.4$) injections included in the O3 injection sets.
If not handled carefully, such holes can cause silent truncation errors.
See the discussion in Appendix~\ref{sec:using the data product}, particularly Appendixes~\ref{sec:detection efficiency} and~\ref{sec:low-dim sampling}.
Users must make sure that their population models do not extend beyond the support of any of the injected distributions used within the cumulative sensitivity estimates when estimating $P(\mathbb{D}|\Lambda)$ with importance sampling.

The software used to execute the injection workflow is publicly available.
\texttt{monte-carlo-vt} \cite{monte-carlo-vt} contains command-line utilities and relies on \texttt{gw-distributions} \cite{gw-distributions} to model probability distributions and \texttt{gw-detectors} \cite{gw-detectors} for IFO responses.
CBC waveform models were accessed via \texttt{lalsimulation} \cite{lalsuite, swiglal}.

The Appendixes provide additional technical details about the injection set and serve as a one-stop-shop with numerical recipes for how to use the public data products.


\section*{Author Contributions}
 
R. Essick authored the software used to generate the injections~\cite{monte-carlo-vt, gw-distributions, gw-detectors}, developed the injected distribution, oversaw the creation of all O4 injections and personally generated all O4a and O4c injections.
He also curated the search results, including preparing LVK's public data releases~\cite{GWTC-4-injections, GWTC-4-cumulative-injections}.
R. Essick wrote the text of the paper, constructed all figures, and performed all calculations within the manuscript.
He developed the one-stop-shop summary of useful calculations presented in Appendix~\ref{sec:using the data product} based on internal LVK technical notes he authored with T. A. Callister and M. Fishbach.
 
M. W. Coughlin, M. Zevin, and D. Chatterjee reviewed the injection software and methodology within the LVK.
M. W. Coughlin and M. Zevin additionally reviewed the public data release.
 
T. A. Clarke, S. Colloms, U. Mali, S. Miller, and N. Steinle generated injections for O4b in consulation with R. Essick.
 
The following authors processed O4a injections with production searches.
Tanmaya Mishra provided results for \texttt{cWB}.
P. Baral, A. C. Baylor, P. Joshi, and C. Messick provided results for \texttt{GstLAL}.
Amazigh Ouzriat provided results for \texttt{MBTA}.
G. C. Davies, T. Dent, P. Kumar, K. S. Phukon, L. Piccari, M. Pillas, and M. Trevor provided results for \texttt{PyCBC}.


\begin{acknowledgments}

R.E. and U.M. are supported by the Natural Sciences \& Engineering Research Council of Canada (NSERC) through a Discovery Grant (RGPIN-2023-03346).
M.W.C. acknowledges support from the National Science Foundation with grant numbers PHY-2308862 and PHY-2117997.
M.Z. gratefully acknowledges funding from the Brinson Foundation in support of astrophysics research at the Adler Planetarium.
D.C. acknowledges support from NSF PHY-2117997.
T.A.C. receives support from the Australian Government Research Training Program.
S.C. is supported by Science and Technology Facilities Council studentship 2748218.
S.J.M. is supported by NSF Grants PHY-2308770 and PHY-2409001.
N.S. acknowledges support from NSERC through the Discovery Grants Program and from the NSERC Canada Research Chairs programs.
A.C.B. is funded by NSF Grant PHY-2207728.
P.B. is funded by NSF Grant PHY-2207728.
P.J. acknowledges support from NSF OAC-2103662.
C.M. acknowledges support from NSF PHY-2207728.
G.C.D. acknowledges the Science and Technology Funding Council (STFC) for funding through grant ST/V005715/1.
K.S.P. acknowledges support from the Science and Technology Funding Council (STFC) grant ST/V005677/1.
M.P. acknowledges support from FNRS and IISN 4.4503.
M.T. acknowledges support from the U.S. National Science Foundation through grant PHY-2409448.

The authors thank Geraint Pratten for his helpful review.

The authors are also grateful for computational resources provided by the LIGO Laboratory and supported by National Science Foundation Grants PHY-0757058 and PHY-0823459.
This material is based upon work supported by NSF's LIGO Laboratory which is a major facility fully funded by the National Science Foundation.

\end{acknowledgments}


\bibliography{biblio}


\appendix
\onecolumngrid


\section{Waveform Approximant}
\label{sec:waveform choices}

For the O4 injection campaign, we modeled CBC signals with the \texttt{IMRPhenomXPHM} approximant \cite{Pratten:2020ceb}, which is fast and natively generated in the frequency domain.
All waveforms were generated with a reference frequency of \result{16 Hz}, and this is the frequency at which the spin components and orbital inclination were defined.

When generating waveforms to estimate $\rho_\mathrm{net,opt}$ for the hopeless cut, we considered frequencies between \result{16 -- 2048 Hz}.
However, when constructing timeseries for searches, we generated waveforms down to \result{10 Hz} and up to \result{8192 Hz} (the Nyquist frequency for a sampling rate of \result{16384 Hz}).
Additionally, to avoid any aliasing while converting the frequency-domain \texttt{IMRPhenomXPHM} waveforms into time-domain data, we first estimated the signal duration ($T$) using a 0-PN approximation for the frequency evolution of the GW's 22-mode
\begin{equation}
    T = \frac{5 \pi^{-8/3}}{256} \left(\frac{G \mathcal{M}^{(\mathrm{det})}}{c^3}\right)^{-5/3} \left(f_\mathrm{low}^{-8/3} - f_\mathrm{high}^{-8/3}\right)
\end{equation}
We then \result{doubled} this estimate, added \result{8 sec}, and then generated waveforms that correspond to \result{durations that are the smallest power of 2 that is greater than ($2T+8$) sec}.
We additionally applied a half-Tukey window (\result{$\alpha = 0.5$}) in the time-domain to control FFT artefacts at the beginning of the time-domain signal. 
This window did not affect any part of the physical signal.

Additionally, during the construction of the O4a injection set, it was discovered that \texttt{IMRPhenomXPHM}'s default multibanding functionality \cite{Pratten:2020ceb} very occasionally produced large, unphysical spikes in the frequency-domain waveform amplitude.
These spikes not only corrupted the time-domain signal; they also produced comically large estimates for $\rho_\mathrm{opt,net}$, sometimes as large as $10^{64}$ or more.
The rate of this error is approximately \result{1 in $3\times10^7$}, which is unfortunately common enough to produce \result{$O(10)$} samples in the O4a injection set.
To avoid this, we turned off multibanding within the waveform calls (see discussion in \citet{Garcia-Quiros:2020qpx}).
While this slowed down waveform generation somewhat, it completely removed all instances of the unphysical amplitudes observed with multibanding.


\section{PSD Variability and Calibration Uncertainty}
\label{sec:psd variability}

We briefly discuss the impact of PSD variability and how we dealt with it during O4 in Sec.~\ref{sec:psd} before discussing calibration uncertainty and its expected impact on sensitivity estimates in Sec.~\ref{sec:cal}.


\subsection{Impact of PSD Variability}
\label{sec:psd}

Fig.~\ref{fig:psd} shows the measured (reference) PSDs for each month of O4a (\result{9 months between May 2023 -- January 2024}, inclusive).
As discussed in Sec.~\ref{sec:estimating psds}, these reference PSDs are the \result{10\%} quantiles in each frequency bin, and the actual noise process may be much more complicated (i.e., nontrivial marginal distributions for the PSD at each frequency and/or correlations between frequency bins).
Nevertheless, variability is clearly visible over $\sim$ month timescales.
It is also worth noting that Fig.~\ref{fig:time} more generally shows that there is non-negligible power in the Fourier transform of $p(t|\mathbb{D},\Lambda)$ for periods of more than a few hours in addition to the clear peaks at one day and one week.

One may be concerned about correctly modeling this variability (and correlations between detectors) in the context of semianalytic sensitivity estimates.
However, for our purposes, the main issue of concern is to make sure that we retain injections (after the hopeless cut) out to large enough redshifts that the distribution detected with a reasonable threshold is never truncated at any point.
\citet{Essick:2023toz} shows that this can effectively be accomplished by adopting a conservative $\rho_\mathrm{cut}$ with a single reference PSD.
Fig.~\ref{fig:time} additionally shows that the amplitude of the variation in $P(t|\mathbb{D},\Lambda)$ is relatively small, almost always less than \result{10\%}.



\subsection{Impact of Calibration Uncertainty}
\label{sec:cal}

Calibration uncertainty in advanced GW IFOs is generally small (see, e.g., Refs.~\cite{Essick:2022vzl, Sun:2020wke, Sun:2021qcg} and references therein).
As such, we do not account for calibration uncertainty with the current injection campaigns.
That is, we inject signals into the data as if the data was perfectly calibrated.

Here, we quantify the size of calibration errors on $P(\mathbb{D}|\theta)$ and $P(\mathbb{D}|\Lambda)$ following the conventions in~\citet{Essick:2023toz}.
Specifically, we consider multiplicative calibration errors in the Fourier domain such that the data contains a distorted version of the true signal ($h$)
\begin{equation}
    d = n + (1+\alpha)e^{i\psi} h
\end{equation}
where $\alpha, \psi \in \mathbb{R}$ may be functions of frequency.
If the true signal is allowed to have an unknown amplitude ($A$) and phase ($\phi$) so that the true signal is $A e^{i\phi} h$, we can analytically maximize the likelihood (assuming stationary Gaussian noise) with respect to $A$ and $\phi$ to obtain a matched-filter response
\begin{equation}
    \rho_\phi^2 = \rho_\mathbb{R}^2 + \rho_\mathbb{I}^2
\end{equation}
where
\begin{align}
    \rho_\mathbb{R} & = \frac{4}{\mathcal{N}} \int df \frac{\mathbb{R}\{d^\ast h\}}{S} \\
    \rho_\mathbb{I} & = \frac{4}{\mathcal{N}} \int df \frac{\mathbb{I}\{d^\ast h\}}{S}
\end{align}
and
\begin{equation}
    \mathcal{N} = \left( 4 \int df \frac{|h|^2}{S} \right)^{1/2}
\end{equation}
$S$ is the one-sided PSD, $d^\ast$ is the complex conjugate of $d$, and $\mathbb{R}\{\cdot\}$ and $\mathbb{I}\{\cdot\}$ are the real and imaginary parts of a complex number, respectively.
It is also common to assume that the template is normalized so that $\mathcal{N} = 1$, but we retain this factor for clarity.
As discussed in~\citet{Essick:2023toz}, for a single IFO, $\rho_\phi$ will be non-central $\chi$-distributed with two degrees of freedom and a centrality parameter $\lambda_\phi$ given by
\begin{align}
    \lambda_\phi^2
        & = \mathrm{E}[\rho_\mathbb{R}]^2 + \mathrm{E}[\rho_\mathbb{I}]^2 \nonumber \\
        & = \rho_\mathrm{opt}^2 \left[ \left(4\int df \frac{|h|^2 (1+\alpha) \cos\psi}{S} \right)^2 + \left(4 \int df \frac{|h|^2 (1+\alpha) \sin\psi}{S} \right)^2 \right] \left(4\int df \frac{|h|^2}{S}\right)^{-2} \label{eq:lambda_phi}
\end{align}
where $\rho_\mathrm{opt} = A \mathcal{N}$ is the optimal signal-to-noise ratio (i.e., the signal-to-noise ratio expected when the signal perfectly matches the template).
Furthermore, it can be shown that $\lambda_\phi = \rho_\mathrm{opt}$ when $\alpha=0$ and $\psi$ is a constant (i.e., no calibration error) and, more generally, that $\lambda_\phi$ is maximized when $\psi$ is a constant.

\begin{figure*}
    \includegraphics[width=1.0\textwidth]{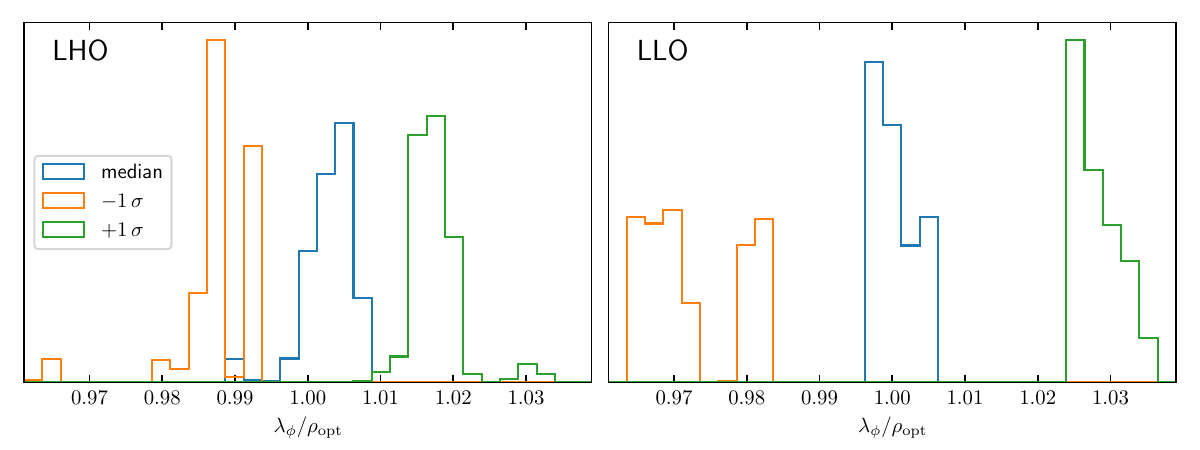}
    \caption{
        Distributions of non-centrality parameter scaled by optimal signal-to-noise ratio in the presence of calibration errors for \result{$> 6000$} hourly calibration snapshots at (\emph{left}) LHO and (\emph{right}) LLO during O3~\cite{O3-Calibration-Envelopes} assuming the reference PSDs from O3 (Fig.~\ref{fig:psd}) and a \result{$30+30\,M_\odot$ (detector-frame) non-spinning binary black hole}.
        Distributions are shown assuming the calibration errors ($\alpha$, $\psi$) follow the (\emph{blue}) median, (\emph{orange}) -1 $\sigma$, and (\emph{green}) +1 $\sigma$ values at all frequencies.
        The $\pm1 \sigma$ bounds are likely overestimates of how large the effects calibration uncertainty would typically be.
    }
    \label{fig:calibration}
\end{figure*}

As shown in~\citet{Essick:2023toz}, the distribution of the observed phase-maximized signal-to-noise ratio $\rho_\phi$ is completely determined by $\lambda_\phi$ and the number of degrees of freedom ($N_\mathrm{dof}$).
This means we can easily compute the survival function
\begin{equation}
    P(\rho_\phi \geq \rho_\mathrm{thr}|\lambda_\phi, N_\mathrm{dof}) = \int\limits_{\rho_\mathrm{thr}}^\infty d\rho_\phi \, p(\rho_\mathrm{\phi}|\lambda_\phi, N_\mathrm{dof})
\end{equation}
which is the basis of the accurate semianalytic sensitivity estimates introduced in~\citet{Essick:2023toz}.
In reality, searches maximize $\rho_\phi$ over a template bank, meaning that estimates only considering the template that matches the true astrophysical signal may, in fact, underestimate the response.

Nevertheless, Fig.~\ref{fig:calibration} shows the distribution of $\lambda_\phi/\rho_\mathrm{opt}$ following Eq.~\ref{eq:lambda_phi} for realistic distributions of calibration errors from \externalresult{$> 6000$} snapshots at each detector during O3~\cite{O3-Calibration-Envelopes}.
It uses the reference PSDs from O3 (Fig.~\ref{fig:psd}) and assumes a \result{$30+30\,M_\odot$ (detector-frame) non-spinning binary black hole}.
The signal-to-noise ratio lost because of calibration errors can be slightly larger for higher masses, but are still typically \result{$\lesssim 5\%$}, from which we determine that the calibration errors observed during O3 were too small to significantly impact sensitivity estimates.

In particular, the median calibration error typically causes a change in the detector response of \result{$\lesssim 1\%$}.
As such, the loss in signal-to-noise associated with calibration errors is very small, comparable to or smaller than the loss in signal-to-noise ratio from the discreteness of template banks.
This loss is small near reasonable detection thresholds ($\rho_\mathrm{thr} \sim 10$) compared to the $O(1)$ variance of $\rho_\phi$, and changing the detection threshold for a semianalytic sensitivity estimate from \result{9.9 to 10.0 or from 10.0 to 10.1 typically changes $P(\mathbb{D}|\Lambda)$ by $\sim 2.5\%$}.
This is comparable to the expected uncertainty in astrophysical rate $\mathcal{K}$ of events for catalogs of \result{$\gtrsim 1600$ events}, much larger than is expected by the end of O4.
However, variability in the calibration error will also generically lead to a broader distribution of $\rho_\phi$, so the effect is not exactly the same as simply changing the detection threshold; see~\citet{Essick:2022vzl} for more discussion.

We also find that difference in $\lambda_\phi$ is almost entirely driven by amplitude calibration ($\alpha$); the phase calibration error ($\psi$) is a significantly smaller contribution.

Finally, even if calibration errors during O4 are a few times worse than during O3, we conclude that these effects can be safely ignored.


\section{Using the Data Product}
\label{sec:using the data product}

We now provide several numerical recipes, their derivations, and example code to compute the detection efficiency and sensitive volume.

It is to be stressed that the sensitive volume and detection efficiency are defined for a distribution (or population) of sources.
As such, all the following expressions are in terms of a ``target'' population that the user must specify.
They also involve integrals over an ``injected'' distribution from which injections are drawn.
These expressions rely on all probability densities (both injected and target distributions) being properly normalized.

It is important to state that (broad) injection sets should \textbf{\textit{not}} be used to compute the
\begin{itemize}
    \item{sensitivity to a particular source (i.e., a $\delta$-function population model) or the}
    \item{sensitivity to sources that were not included within the support of the injected distribution.}
\end{itemize}
The first will result in a very noisy Monte Carlo approximation to integrals because the target distribution (a $\delta$-function) is very different from a broad injection distribution.
The second will result in a (silent) truncation error.
That is, there may be no obvious failure, but the integral will be misestimated because the support of the injected distribution does not completely span the support of the target distribution, thereby acting as an additional artificial selection.


\subsection{Detection Efficiency}
\label{sec:detection efficiency}

We define the detection efficiency, or the probability of detection $P(\mathbb{D}|\Lambda)$ (i.e., the fraction of a population that would be detected), in Eq.~\ref{eq:P(D|Lambda)}.
The average is taken over detector-frame time, a population of sources, and noise realizations within the detectors.
We can express this as the ratio of the expected number of detections ($N_\mathrm{det}$) and the total number of binaries ($\mathcal{K}$) that occur within the past light-cone of an observation period: $\mathrm{E}[N_\mathrm{det}] = \mathcal{K} P(\mathbb{D}|\Lambda)$.

Some analysts may wish to estimate $P(\mathbb{D}|\Lambda)$ for each of a set of small populations (e.g., bins in a histogram) by drawing many injections from a broad injection distribution and directly counting $N_\mathrm{det}$ and $N_\mathrm{fnd}$ within each bin.
However, this can require a computationally infeasible number of draws to estimate $N_\mathrm{fnd}$ for nearby (low-$z$) populations if the draw distribution is broad (i.e., it requires tracking the parameters of too many samples ``without a hopeless cut'').
We will show how to instead estimate this using only lists of found injections, their associated draw probabilities, and the total number of injections drawn from $p(\theta|\Lambda)$.

In general, one can approximate an integral with the sum over Monte Carlo samples via
\begin{equation}
    \int dx \, p(x) f(x) \approx \frac{1}{M} \sum\limits_e^M f(x_e)
\end{equation}
where the $M$ samples are i.i.d. from $p(x)$
\begin{equation}
    x_e \sim p(x) \ \forall \ x_e
\end{equation}
What's more, if we wish to approximate $\int dx\,q(x)f(x)$ but only have $M$ samples $\{x_e\}$ drawn from $p(x)$, we can do so by noting that
\begin{align}
    \int dx \, q(x) f(x)
        & = \int dx \, p(x) \left[ \frac{q(x)}{p(x)} f(x) \right] \nonumber \\
        & \approx \frac{1}{M} \sum\limits_e^M \left[\frac{q(x_e)}{p(x_e)} f(x_e)\right]
\end{align}
This is referred to as importance sampling, and it can be done for any target distribution $q(x)$ as long as the support of $q(x)$ is entirely covered by the support of the draw distribution $p(x)$.

Let's use importance sampling to estimate $P(\mathbb{D}|\Lambda)$.
Eq.~\ref{eq:P(D|Lambda) short} states
\begin{equation} \nonumber
    \hat{P}(\mathbb{D}|\Lambda) \equiv \frac{1}{N_\mathrm{inj}} \sum\limits_e^{N_\mathrm{fnd}} \frac{p(\theta_e|\Lambda)}{p(\theta_e|\Lambda_\mathrm{inj})}
\end{equation}
The requirements for this approach are that
\begin{itemize}
    \item $p(\theta|\Lambda_\mathrm{inj})$ must have support everywhere that $p(\theta|\Lambda)$ does,
    \item both $p(\theta|\Lambda)$ and $p(\theta|\Lambda_\mathrm{inj})$ must be properly normalized, and
    \item both $p(\theta|\Lambda)$ and $p(\theta|\Lambda_\mathrm{inj})$ must be defined over the same variables.
\end{itemize}
For example, if $p(\theta|\Lambda)$ is convenient to express in terms of polar spin coordinates but $p(\theta|\Lambda_\mathrm{inj})$ is specified in terms of Cartesian spin components, then one must include a Jacobian to convert one of the distributions to match the other.
Similarly, if $p(\theta|\Lambda_\mathrm{inj})$ is specified over component masses, redshifts, and spins, then $p(\theta|\Lambda)$ must also be specified over component masses, redshifts, and spins.

If $p(\theta|\Lambda_\mathrm{inj})$ is separable (i.e., $p(z, m_1, m_2, \vec{s}_1, \vec{s}_2|\Lambda_\mathrm{inj}) = p(z|\Lambda_\mathrm{inj}) p(m_1, m_2|\Lambda_\mathrm{inj}) p(\vec{s}_1, \vec{s}_2|\Lambda_\mathrm{inj})$) and the separate draw probabilities are individually available, then one can adopt parts of the draw probability within the target distribution, thereby specifying a new target distribution for only some variables.
That is, if the draw probability is separable and the target distribution is
\begin{equation}
    p(z, m_1, m_2, s_1, s_2|\Lambda) = p(m_1, m_2|\Lambda) p(z|\Lambda_\mathrm{inj}) p(s_1, s_2|\Lambda_\mathrm{inj})
\end{equation}
then
\begin{align}
    \hat{P}(\mathbb{D}|\Lambda)
        & = \frac{1}{N_\mathrm{inj}}\sum\limits_e^{N_\mathrm{fnd}} \frac{p(m_{1,e}, m_{2,e}|\Lambda) p(z_e|\Lambda_\mathrm{inj}) p(s_{1,e}, s_{2,e}|\Lambda_\mathrm{inj})}{(m_{1,e}, m_{2,e}|\Lambda_\mathrm{inj}) p(z_e|\Lambda_\mathrm{inj}) p(s_{1,e}, s_{2,e}|\Lambda_\mathrm{inj})} \nonumber \\
        & = \frac{1}{N_\mathrm{inj}}\sum\limits_e^{N_\mathrm{fnd}} \frac{p(m_{1,e}, m_{2,e}|\Lambda)}{p(m_{1,e}, m_{2,e}|\Lambda_\mathrm{inj})} \label{eq:shortcut}
\end{align}

As a concrete example, let's compute $\hat{P}(\mathbb{D}|\Lambda)$ for separate bins in a histogram over $m_1$.
For any given bin $\alpha$ bounded by $\mathrm{min}_\alpha$ and $\mathrm{max}_\alpha$, the associated target distribution could be
\begin{equation}
    p(m_1|\Lambda_\alpha) = \frac{\Theta(\mathrm{min}_\alpha \leq m_1 < \mathrm{max}_\alpha)}{\mathrm{max}_\alpha - \mathrm{min}_\alpha}
\end{equation}
Inserting this into Eq.~\ref{eq:shortcut} and repeating for all bins provides a direct way to estimate the detection efficiency across parameter space.
This is entirely equivalent to a ``normal'' histogram but does not require knowledge of anything beyond the set of found injections, their draw probabilities, and the total number of injections drawn.


\subsubsection{Example Implementation}

With modern file formats, it is easily to construct these types of Monte Carlo sums.
Below is an example of how one might compute the detection efficiency for a very simple population over (source-frame) component masses while adopting the injected distribution for the rest of the parameters.

\begin{verbatim}
import numpy
import h5py

#---

# declare the path to tabular data
path = "path/to/samples.hdf"

#---

# load the required data
with h5py.File(path, "r") as obj:
    events = obj["events"][:]
    meta = dict(obj.attrs.items())

#---

# extract weights for mixture model
# see Essick, Research Notes of the AAS 5, 220 (2021).
weights = events["weights"]

#---

# extract p(m1, m2|draw)
pdraw = numpy.exp(events["lnpdraw_mass1_source"] \
    + events["lnpdraw_mass2_source_GIVEN_mass1_source"])

# compute target distribution: flat in component masses between 10 - 20 Msun
# Note that we must include the proper normalization.
m1 = events["mass1_source"]
m2 = events["mass2_source"]

ptarget = (10 <= m1)*(m1 < 20) * (10 <= m2)*(m2 < 20) / (20 - 10)**2 / 2

#---

### identify which events were detected using a FAR threshold
min_far = numpy.min([events["%s_far"%search] for search in meta["searches"]], axis=0)
detected = min_far < 1.0 # /year

### compute the Monte Carlo sum
pdet = numpy.sum(weights[detected] * ptarget[detected] / pdraw[detected]) / meta["total_generated"]
\end{verbatim}


\subsection{Sensitive Volume}

The sensitive volume, often denoted $\left<V\right>$, is related to $P(\mathbb{D}|\Lambda)$.
However, it is associated with specific assumptions about the population model.
That is, it assumes the sources are uniformly distributed in comoving volume and source-frame time:
\begin{equation}\label{eq:uniform in co-moving volume}
    \frac{dN}{dV_c dt_\mathrm{src} d\phi} = \mathcal{R}_{V_c,\mathrm{src}} \,  p(\phi|\Lambda)
\end{equation}
where $\mathcal{R}_{V_c,\mathrm{src}}$ represents the rate per comoving volume per source-frame time and $\phi$ represents the variables in $\theta$ besides $z$.
This implies
\begin{align}
    \frac{dN}{dt_\mathrm{det} dz d\phi}
        & = \mathcal{R}_{V_c,\mathrm{src}} \left( \frac{dV_c}{dz} \frac{dt_\mathrm{src}}{dt_\mathrm{det}} \right) p(\phi|\Lambda) \nonumber \\
        & = \mathcal{R}_{V_c,\mathrm{src}} \left( \frac{dV_c}{dz} \frac{1}{1+z} \right) p(\phi|\Lambda) \label{eq: local merger rate}
\end{align}
Another way to say this is that $p(z,\phi|\Lambda)$ is separable and that $p(z|\Lambda) \propto (dV_c/dz) (1+z)^{-1}$.
Here, the distance to the source is represented by either the redshift ($z$) or the contained comoving volume ($V_c$).

The sensitive volume is then defined as an average over $p(\phi|\Lambda)$
\begin{equation}
    \left<V\right> = \int dz \, \frac{dV_c}{dz} \frac{1}{1+z} \int d\phi\, p(\phi|\Lambda) P(\mathbb{D}|z, \phi)
\end{equation}
so that
\begin{equation}
    N_\mathrm{det} = \mathcal{R}_{V_c,\mathrm{src}} \left<V\right> T \label{eq:simple rate}
\end{equation}
given the model in Eq.~\ref{eq:uniform in co-moving volume} and the detector-frame survey duration $T$.
Eq.~\ref{eq:simple rate} can be used to estimate $\mathcal{R}_{V_c,\mathrm{src}}$ given $N_\mathrm{det}$, although one can also estimate the rate for more general models which do not assume the rate is constant (Sec.~\ref{sec:rate}).

The integral expression for $\left<V\right>$ can also be approximated via importance sampling
\begin{align}
    \left<V\right>
        & = \int d\phi\, p(\phi|\Lambda) \int dz \frac{dV_c}{dz}\frac{1}{1+z} P(\mathbb{D}|z, \phi) \nonumber \\
        & \approx \frac{1}{N_\mathrm{inj}} \sum\limits_e^{N_\mathrm{fnd}} \left[ \frac{(dV_c/dz)|_{z_e} (1+z_e)^{-1} p(\phi_e|\Lambda)}{p(z_e, \phi_e|\Lambda_\mathrm{inj})} \right] \label{eq:Vhat}
\end{align}

Again, the usual criteria apply for the support of $p(\phi|\Lambda)$ and $p(\phi|\Lambda_\mathrm{inj})$.
We also need $p(z|\Lambda_\mathrm{inj})$ to extend to high enough $z$ that $P(\mathbb{D}|z) \sim 0$ in order to avoid truncation errors.
Finally, if $p(z,\phi|\Lambda_\mathrm{inj})$ is separable, analysts may again adopt parts of the draw probability into the target distribution and further simplify the sum.

Note that most injection sets are drawn uniformly in detector-frame wall-time, meaning some injections occur at times when the detectors are out of science mode.
The sum over found injections therefore automatically includes the effects of the detectors' duty cycles, and one should multiply Eq.~\ref{eq:Vhat} by the total detector-frame wall-time (and not the total science time) to estimate $\left<V\right>T$.


\subsection{Local Merger Rate}
\label{sec:rate}

One can more generally estimate the local merger rate ($R_{V_c,\mathrm{src}}$) at a particular redshift even if the local rate is not constant.
We begin by generalizing Eq.~\ref{eq: local merger rate} to allow for a local merger rate that depends on redshift.
\begin{equation}
    \frac{dN}{dt_\mathrm{det} dz d\phi} = \mathcal{R}_{V_c,\mathrm{src}}(z|\Lambda) \left(\frac{dV_c}{dz} \frac{1}{1+z}\right) p(\phi|\Lambda)
\end{equation}
From this, we can immediately compute the expected number of detections
\begin{align}
    \mathrm{E}[N_\mathrm{det}|T, \Lambda]
        & = \int dt_\mathrm{det} dz d\phi \, \mathcal{R}_{V_c,\mathrm{src}}(z|\Lambda) \left(\frac{dV_c}{dz}\frac{1}{1+z}\right) p(\phi|\Lambda) P(\mathbb{D}|t_\mathrm{det}, z, \phi) \nonumber \\
        & = T \left< \int dz d\phi \, \mathcal{R}_{V_c,\mathrm{src}}(z|\Lambda) \left(\frac{dV_c}{dz}\frac{1}{1+z}\right) p(\phi|\Lambda) P(\mathbb{D}|t_\mathrm{det}, z, \phi) \right>_{t_\mathrm{det}} \nonumber \\
        & \approx T \left( \frac{1}{N_\mathrm{inj}} \sum\limits_e^{N_\mathrm{fnd}} \frac{\mathcal{R}_{V_c,\mathrm{src}}(z_e|\Lambda) (dV_c/dz)|_{z_e} (1+z_e)^{-1} p(\phi_e|\Lambda)}{p(z_e, \phi_e|\Lambda_\mathrm{inj})} \right) \label{eq:expected number of detections}
\end{align}
where $\left< \cdot \right>_{t_\mathrm{det}}$ in the second line refers to an average over the duration of the experiment.
What's more, we can write the hierarchical posterior (Eq.~\ref{eq:joint}) as
\begin{align}
    p(\mathcal{K}, \Lambda| \{d_e\}, N_\mathrm{det})
        & \sim p(\mathcal{K}, \Lambda) \mathcal{K}^{N_\mathrm{det}} e^{-\mathcal{K}P(\mathbb{D}|\Lambda)} \prod\limits_e^{N_\mathrm{det}} \int d\theta p(d_e|\theta) p(\theta|\Lambda) \nonumber \\
        & \sim \left[ p(\mathcal{K}) \left(\mathcal{K}P(\mathbb{D}|\Lambda)\right)^{N_\mathrm{det}} e^{-\mathcal{K} P(\mathbb{D}|\Lambda)} \right] \left[ p(\Lambda) \prod\limits_e^{N_\mathrm{det}} \frac{\int d\theta p(d_e|\theta) p(\theta|\Lambda)}{P(\mathbb{D}|\Lambda)} \right]
\end{align}
assuming $p(\mathcal{K}, \Lambda) = p(\mathcal{K}) p(\Lambda)$.
If we instead place a prior on the expected number of detections rather than the number of astrophysical events ($\mathrm{E} = \mathcal{K} P(\mathbb{D}|\Lambda)$; see~\citet{Essick:2023upv}), we can further simplify this to
\begin{equation}
    p(\mathrm{E}, \Lambda|\{d_e\}, N_\mathrm{det}) = p(\mathrm{E}|N_\mathrm{det}) p(\Lambda|\{d_e\}, N_\mathrm{det})
\end{equation}
with
\begin{align}
    p(\Lambda|\{d_e\}, N_\mathrm{det}) & \sim p(\Lambda) \prod\limits_e^{N_\mathrm{det}} \frac{\int d\theta p(d_e|\theta) p(\theta|\Lambda)}{P(\mathbb{D}|\Lambda)} \label{eq:Lambda post} \\
    p(\mathrm{E}|N_\mathrm{det}) & \sim p(\mathrm{E}) \mathrm{E}^{N_\mathrm{det}} e^{-\mathrm{E}}
\end{align}
This is commonly done by assuming $p(\mathcal{K}) \sim 1/\mathcal{K}$ and therefore $p(\mathrm{E}) \sim 1/\mathrm{E}$, in which case $\mathrm{E}|N_\mathrm{det}$ is $\Gamma$-distributed
\begin{equation} \label{eq:E post}
    p(\mathrm{E} | N_\mathrm{det}) = \frac{\mathrm{E}^{N_\mathrm{det}-1}}{\Gamma(N_\mathrm{det})} e^{-\mathrm{E}}
\end{equation}
By combining Eqs.~\ref{eq:expected number of detections} and~\ref{eq:E post}, it is straightforward to show that the posterior for the rate at a specific redshift (conditioned on a particular population) can be written as
\begin{align}
    p(\mathcal{R}_{V_c,\mathrm{src}}(z) | N_\mathrm{det}, \Lambda, T)
        & = p(\mathrm{E}|N_\mathrm{det}) \frac{d\mathrm{E}}{d\mathcal{R}_{V_c,\mathrm{src}}} \nonumber \\
        & = p(\mathrm{E}|N_\mathrm{det}) \frac{\mathrm{E}}{\mathcal{R}_{V_c,\mathrm{src}}} \nonumber \\
        & = \left( \frac{\mathrm{E}^{N_\mathrm{det}} e^{-\mathrm{E}}}{\Gamma(N_\mathrm{det})} \right) \frac{1}{\mathcal{R}_{V_c,\mathrm{src}}(z|\Lambda)} \label{eq:post rate}
\end{align}
where $\mathrm{E}$ is an implicit function of $\mathcal{R}_{V_c,\mathrm{src}}$, $T$, and $\Lambda$.
Marginalization over $\Lambda$ can then be achieved by taking the average of Eq.~\ref{eq:post rate} with respect to samples drawn from the corresponding hyperposterior (Eq.~\ref{eq:Lambda post}).

Given a set of hyperposterior samples for $\Lambda$, the following yields samples from the marginal distribution for $\mathcal{R}_{V_c,\mathrm{src}}(z)$.
For convenience, we will write $\mathcal{R}_{V_c,\mathrm{src}}(z) = \mathcal{R}_o f(z)$ where $f(z=0)=1$ and therefore $\mathcal{R}_o$ represents the local merger rate at $z=0$.
The procedure is then
\begin{enumerate}
    \item draw $\mathrm{E} \sim \mathrm{E}^{N_\mathrm{det}-1} e^{-\mathrm{E}} / \Gamma(N_\mathrm{det})$
    \item draw $\Lambda \sim p(\Lambda|\{d_e\}, N_\mathrm{det})$
    \item compute $\mathcal{R}_o = (\mathrm{E} / T) \left( N_\mathrm{inj}^{-1} \sum_e^{N_\mathrm{fnd}} \left[ (dV_c/dz)|_{z_e}(1+z_e)^{-1} f(z_e|\Lambda) p(\phi_e|\Lambda) / p(z_e,\phi_e|\Lambda_\mathrm{inj}) \right] \right)^{-1}$
    \item compute $\mathcal{R}_{V_c,\mathrm{src}} = \mathcal{R}_o f(z|\Lambda)$
\end{enumerate}
This can be repeated until a representative set is obtained.


\subsection{Estimating Selection with Alternate Cosmologies}

We can still perform exactly the same integrals for $P(\mathbb{D}|\Lambda)$ and $\left<V\right>$ with a different cosmology than was assumed within $p(\theta|\Lambda_\mathrm{inj})$.
Specifically, by changing coordinates from, e.g., source-frame component masses and redshift to detector-frame component masses and luminosity distance, we can compute the selection entirely in the detector frame and therefore for any expansion history $D_L(z)$.
To wit, we write
\begin{align}
    P(\mathbb{D}|\Lambda)
        & = \frac{1}{N_\mathrm{inj}} \sum\limits_e^{N_\mathrm{fnd}} \frac{p(m_{1,e}^{(\mathrm{det})}, m_{2,e}^{(\mathrm{det})}, D_{L,e}|\Lambda)}{p(m_{1,e}^{(\mathrm{det})}, m_{2,e}^{(\mathrm{det})}, D_{L,e}|\Lambda_\mathrm{inj})} \nonumber \\
        & = \frac{1}{N_\mathrm{inj}} \sum\limits_e^{N_\mathrm{fnd}} \frac{|dD_L/dz|_{z_e}^{-1} (1+z_e)^{-2} p(m_{1,e}^{(\mathrm{src})}=m_{1,e}^{(\mathrm{det})}/(1+z_e), m_{2,e}^{(\mathrm{src})}=m_{2,e}^{(\mathrm{det})}/(1+z_e), z_e|\Lambda)}{|dD_L^\prime/dz^\prime|_{z^\prime_e}^{-1} (1+z^\prime_e)^{-2} p(m_{1,e}^{(\mathrm{src})}=m_{1,e}^{(\mathrm{det})}/(1+z^\prime_e), m_{2,e}^{(\mathrm{src})}=m_{2,e}^{(\mathrm{det})}/(1+z^\prime_e), z^\prime_e|\Lambda_\mathrm{inj})}
\end{align}
where $D_L$ and $z$ correspond to the target cosmology and $D_L^\prime$ and $z^\prime$ correspond to the injected cosmology.
See, e.g.,~\citet{Gray:2023wgj} for more discussion.

The public data products~\cite{GWTC-4-cumulative-injections, GWTC-4-injections} record $D_L^\prime$, $z^\prime$, and $dD_L^\prime/dz^\prime$ for each injection alongside the injected probability density $p(\theta|\Lambda_\mathrm{inj})$.


\subsection{Importance Sampling over Lower-Dimensional Spaces}
\label{sec:low-dim sampling}

GWs from CBCs are often parametrized in terms of Cartesian spin components for each object at a reference frequency.
For this reason, injections (posterior samples) are often specified in terms of known injected (prior) distributions over the spin components.
However, it can be useful to instead fit the distribution of effective spins within a population instead of the 6D spin-component distributions.
Because the effective spins span a lower dimensional space than the spin components, it may not be straightforward to compute the induced marginal injected (prior) distribution over the effective spins for arbitrary distributions over the 6D spin components, thereby stymieing typical importance sampling sums like Eq.~\ref{eq:Phat(D|Lambda)}.

We demonstrate a convenient regrouping of terms within the importance sampling summand to resolve this issue.
That is, as long as one knows the mapping from at least one 6D component spin distribution (with support throughout the entire physically allowed parameter space) to the induced distribution over effective spins, then one can exploit this to compute importance sampling sums over effective spin distributions starting from arbitrary distributions over the 6D spin components.

Let us begin by asserting that we are interested in estimating integrals like
\begin{equation}
    F = \int d\chi \, p(\chi|\Lambda) f(\chi)
\end{equation}
based on a set of samples $\{\vec{s}_{1}, \vec{s}_2\}$ drawn from $p(\vec{s}_1, \vec{s}_2|\Lambda_\mathrm{inj})$ where $\chi = \chi(\vec{s}_1, \vec{s}_2)$.
Typically, we might approach this with Monte Carlo importance sampling and approximate
\begin{align}
    F & = \int d\chi\, p(\chi|\Lambda_\mathrm{inj}) \left[ f(\chi) \frac{p(\chi|\Lambda)}{p(\chi|\Lambda_\mathrm{inj})} \right] \nonumber \\
      & \approx \frac{1}{M} \sum\limits_{a}^M f(\chi_a) \frac{p(\chi_a|\Lambda)}{p(\chi_a|\Lambda_\mathrm{inj})} \label{eq:2}
\end{align}
where $\chi_a \sim p(\chi|\Lambda_\mathrm{inj})$.
However, as we do not know how to compute $p(\chi|\Lambda_\mathrm{inj})$ easily from a general distribution $p(\vec{s}_1,\vec{s}_2|\Lambda_\mathrm{inj})$, this approach is not (immediately) profitable.

The key insight is that we do know how to compute $p(\chi|\Lambda)$ from $p(\vec{s}_1,\vec{s}_2|\Lambda)$ for at least a few special cases with broad support.
Specifically, this mapping is known when the spins for each object that are i.i.d. from isotropically distributed orientations with uniformly distributed magnitudes; see \citet{Callister:2021gxf} or \citet{Iwaya:2024zzq} for explicit derivations.
We refer to this distribution as $p(\cdot|\Lambda_\mathrm{iso})$.
We also know how to reweight samples in the full 6D space to appear as though they have been drawn from different distributions.
We combine these as follows.
\begin{align}
    F & = \int d\chi\, p(\chi|\Lambda) f(\chi) \nonumber \\
      & = \int d\chi\, p(\chi|\Lambda_\mathrm{iso}) \left[ f(\chi) \frac{p(\chi|\Lambda)}{p(\chi|\Lambda_\mathrm{iso})} \right] \nonumber \\
      & = \int d\chi\, \left[ \int d\vec{s}_1 d\vec{s}_2\, p(\vec{s}_1, \vec{s}_2|\Lambda_\mathrm{iso}) \delta(\chi - \chi(\vec{s}_1, \vec{s}_2)) \right]  \left[ f(\chi) \frac{p(\chi|\Lambda)}{p(\chi|\Lambda_\mathrm{iso})} \right] \nonumber \\
      & = \int d\vec{s}_1 d\vec{s}_2\, p(\vec{s}_1, \vec{s}_2|\Lambda_\mathrm{iso}) \left[ f(\chi=\chi(\vec{s}_1, \vec{s}_2)) \frac{p(\chi=\chi(\vec{s}_1, \vec{s}_2)|\Lambda)}{p(\chi=\chi(\vec{s}_1, \vec{s}_2)|\Lambda_\mathrm{iso})} \right]
\end{align}
Up to this point, we have followed the approach in Eq.~\ref{eq:2} but also exploited the fact that the measure $d\chi \, p(\chi|\Lambda_\mathrm{iso})$ can be derived as an integral over $d\vec{s}_1 d\vec{s}_2\, p(\vec{s}_1, \vec{s}_2|\Lambda_\mathrm{iso})$, sometimes referred to as the law of the unconscious statistician.
As a reminder, this approach requires the support of $p(\chi|\Lambda)$ to be contained with the support of $p(\chi|\Lambda_\mathrm{iso})$.

We additionally know how to reweight samples in the 6D spin space, and doing so yields
\begin{align}
    F & = \int d\vec{s}_1 d\vec{s}_2\, p(\vec{s}_1, \vec{s}_2|\Lambda_\mathrm{inj}) \left[ \frac{p(\vec{s}_1, \vec{s}_2|\Lambda_\mathrm{iso})}{p(\vec{s}_1, \vec{s}_2|\Lambda_\mathrm{inj})} \right] \left[ f(\chi=\chi(\vec{s}_1, \vec{s}_2)) \frac{p(\chi=\chi(\vec{s}_1, \vec{s}_2)|\Lambda)}{p(\chi=\chi(\vec{s}_1, \vec{s}_2)|\Lambda_\mathrm{iso})} \right]
\end{align}
which requires the support of $p(\vec{s}_1, \vec{s}_2|\Lambda_\mathrm{iso})$ to be entirely contained within the support of $p(\vec{s}_1, \vec{s}_2|\Lambda_\mathrm{inj})$.
Finally, we can approximate the integral over $\vec{s}_1$ and $\vec{s}_2$ with samples drawn from $p(\vec{s}_1, \vec{s}_2|\Lambda_\mathrm{inj})$ via
\begin{align}
    F & \approx \frac{1}{M} \sum\limits_e^M f(\chi_e) \frac{p(\chi_e|\Lambda)}{w_e} \label{eq:this one}
\end{align}
where
\begin{align}
    \vec{s}_{1e}, \vec{s}_{2e} & \sim p(\vec{s}_1, \vec{s}_2|\Lambda_\mathrm{inj})\\
    \chi_e & = \chi(\vec{s}_{1e}, \vec{s}_{2e}) \\
    w_e & = p(\chi=\chi_e|\Lambda_\mathrm{iso}) \left[ \frac{p(\vec{s}_{1e},\vec{s}_{2e}|\Lambda_\mathrm{inj})}{p(\vec{s}_{1e}, \vec{s}_{2e}|\Lambda_\mathrm{iso})} \right]
\end{align}
Again, this derivation requires the support of $p(\chi|\Lambda)$ to be entirely within the support of $p(\chi|\Lambda_\mathrm{iso})$ and the support of $p(\vec{s}_1, \vec{s}_2|\Lambda_\mathrm{iso})$ to be entirely contained within the support of $p(\vec{s}_1, \vec{s}_2|\Lambda_\mathrm{inj})$.

Importantly, if injected distributions have mass-dependent spin distributions (like those produced for O3 \cite{GWTC-3-injections, GWTC-3-cumulative-injections}), then separable target distributions over both mass and spin (i.e., $p(m, s) = p(m) p(s)$) and $p(s|\Lambda_\mathrm{iso})$, used as an intermediate step, must limit their spins to the smallest domain of $p(s|m,\Lambda_\mathrm{inj})$ for all $m$.

\end{document}